\documentclass[lettersize,journal]{IEEEtran}
\usepackage{amsmath,amsfonts}
\usepackage{algorithmic}
\usepackage{algorithm}
\usepackage{array}
\usepackage[caption=false,font=normalsize,labelfont=sf,textfont=sf]{subfig}
\usepackage{textcomp}
\usepackage{stfloats}
\usepackage{url}
\usepackage{verbatim}
\usepackage{graphicx}
\usepackage{cite}
\usepackage{glossaries}
\usepackage{caption}
\usepackage{blindtext}
\usepackage[dvipsnames]{xcolor}
\usepackage[hidelinks]{hyperref}
\usepackage{orcidlink}
\usepackage[noabbrev,capitalise]{cleveref}
\usepackage{booktabs}
\usepackage{adjustbox}
\usepackage{siunitx}
\usepackage[switch]{lineno}
\usepackage{xurl}
\usepackage{orcidlink}
\usepackage{tikz}

\newcommand*\circled[1]{\tikz[baseline=(char.base)]{
            \node[shape=circle,draw,inner sep=2pt] (char) {#1};}}

\begin{document}

\newacronym{ai}{AI}{Artificial Intelligence}
\newacronym{ml}{ML}{Machine Learning}
\newacronym{nn}{NN}{Neural Network}
\newacronym{prb}{PRB}{Physical Resource Block}
\newacronym{tti}{TTI}{Transmission Time Interval}
\newacronym{ppa}{PPA}{Power Performance and Area}

\newacronym{cpu}{CPU}{Central Processing Unit}
\newacronym{asic}{ASIC}{Application Specific Integrated Circuit}
\newacronym{fpga}{FPGA}{Field Programmable Gate Array}
\newacronym{gpu}{GPU}{Graphics Processing Unit}

\newacronym{6g}{6G}{6th Generation}
\newacronym{gnb}{gNB}{Next Generation Node B}
\newacronym{phy}{PHY}{Physical Layer}
\newacronym{ofdma}{OFDMA}{Orthogonal Frequency Division Multiple Access}
\newacronym{ber}{BER}{Bit-Error-Rate}
\newacronym{lmmse}{LMMSE}{Linear Minimum Mean Squared Error}
\newacronym{ran}{RAN}{Radio Access Network}
\newacronym{pe}{PE}{Processing Element}
\newacronym{te}{TE}{Tensor Engine}
\newacronym{gemm_formula}{GEMM, $Z=Y+(X*W)$}{General Matrix-Matrix Multiply}
\newacronym{gemm}{GEMM}{General Matrix-Matrix Multiply}
\newacronym{fma}{FMA}{FP16-Multiply\&Add unit}

\newacronym{rob}{ROB}{Reorder Buffer}
\newacronym{fifo}{FIFO}{First-In First-Out}
\newacronym{xbar}{XBAR}{crossbar}
\newacronym{cfft}{CFFT}{Complex Fast Fourier Transform}
\newacronym{fpu}{FPU}{Floating-Point Unit}
\newacronym{rtl}{RTL}{Register Transfer Level}
\newacronym{dma}{DMA}{Direct Memory Access}

\newacronym{mol}{MoL}{Memory on Logic}
\newacronym{beol}{BEOL}{Back End of the Line}
\newacronym{bsm}{BSM}{Back Side Metal}
\newacronym{npu}{NPU}{Neural Processing Unit}
\newacronym{hbm}{HBM}{High Bandwidth Memory}

\newacronym{ipc}{IPC}{Instructions Per Cycle}
\newacronym{ls}{LS}{Least Squares}
\newacronym{mmse}{MMSE}{Minimum Mean Squared Error}
\newacronym{che}{CHE}{Channel Estimation}
\newacronym{mimo}{MIMO}{Multi-Input Multi-Output}
\newacronym{re}{RE}{Resource Element}
\newacronym{prb}{PRB}{Physical Resource Block}
\newacronym{FP16}{FP16}{16-bit Floating-Point}

\newacronym{mha}{MHA}{Multi-Head Attention}
\newacronym{relu}{ReLU}{Rectified Linear Unit}
\newacronym{resnet}{ResNet}{Residual Network}

\newacronym{fc}{FC}{Fully-Connected}
\newacronym{sm}{SM}{Streaming Multiprocessor}
\newacronym{asip}{ASIP}{Application Specific Instruction Processor}

\title{TensorPool: A 3D-Stacked 8.4TFLOPS/4.3W Many-Core Domain-Specific Processor for AI-Native Radio Access Networks}


\author{Marco~Bertuletti \orcidlink{0000-0001-7576-0803},
        Yichao Zhang \orcidlink{0009-0008-7508-599X},
        Diyou Shen \orcidlink{0009-0008-8014-7948},
        Alessandro~Vanelli-Coralli \orcidlink{0000-0002-4475-5718},
        Frank~K.~G{\"u}rkaynak \orcidlink{0000-0002-8476-554X},
        Luca~Benini \orcidlink{0000-0001-8068-3806}
\thanks{Marco Bertuletti, Yichao Zhang, Diyou Shen, and Frank~K.~G{\"u}rkaynak are with the Integrated Systems Laboratory (IIS), Eidegenossische Technische Hochschule (ETH), Zurich, Switzerland e-mail: mbertuletti@iis.ee.ethz.ch, yiczhang@iis.ee.ethz.ch, dishen@iis.ee.ethz.ch, kgf@iis.ee.ethz.ch. Alessandro Vanelli-Coralli and Luca Benini are with the University of Bologna, Bologna, Italy, and with ETH e-mail: avanelli@iis.ee.ethz.ch, lbenini@iis.ee.ethz.ch.}
}



\maketitle

\begin{abstract}
 The upcoming integration of AI in the physical layer (PHY) of 6G radio access networks (RAN) will enable a higher quality of service in challenging transmission scenarios.
 However, deeply optimized AI-Native PHY models impose higher computational complexity compared to conventional baseband, challenging deployment under the sub-msec real-time constraints typical of modern PHYs. Additionally, following the extension to terahertz carriers, the upcoming densification of 6G cell-sites further limits the power consumption of base-stations, constraining the budget available for compute ($\leq$~100~W).
 The desired flexibility to ensure long term sustainability and the imperative energy-efficiency gains on the high-throughput tensor computations dominating AI-Native PHYs can be achieved by domain-specialization of many-core programmable baseband processors.
 Following the domain-specialization strategy, we present TensorPool, a cluster of 256 RISCV32IMAF programmable cores, accelerated by 16 256~MACs/cycle (FP16) tensor engines with low-latency access to 4MiB of L1 scratchpad for maximal data-reuse. Implemented in TSMC's N7, TensorPool achieves 3643~MACs/cycle (89\% tensor-unit utilization) on tensor operations for AI-RAN, 6$\times$ more than a core-only cluster without tensor acceleration, while simultaneously improving GOPS/W/mm\textsuperscript{2} efficiency by 9.1$\times$. Further, we show that 3D-stacking the computing blocks of TensorPool to better unfold the tensor engines to L1-memory routing provides 2.32$\times$ footprint improvement with no frequency degradation, compared to a 2D implementation.
\end{abstract}


\section{Introduction}

The IMT-2030 recommendation for the next-generation \glspl{ran}~\cite{itu_2030} promotes the dense integration of \gls{ai} and communication, with two main advantages. First, a \gls{ml} based \gls{ran} is expected to support self-monitoring, self-organization, self-optimization, and self-healing under various traffic conditions. Second, empowering the \gls{ran} infrastructure with hardware for \gls{ml} accelerated computing will create a distributed computer for the \gls{ai} computing-continuum~\cite{li_computingcommunications_2023}. This will ensure seamless user experiences across devices from the edge to the cloud and enable distributed intelligent and data-driven applications.

Focusing on the first objective, on-field studies demonstrated that \gls{ai} has an edge over model-driven approaches~\cite{li_airanoperators_2025}: it can learn scenarios where explicit modeling is infeasible, it can make real-time decisions in high-dimensional design-spaces, and it can perform accurate prediction and proactive optimization. These are valuable capabilities in the \gls{phy}, where \gls{ml} based solutions were applied to channel estimation, beamforming, dynamic power adjustment, and interference mitigation, but also in the data link layer and the network layer, where \gls{ai} solutions were devised for adaptive modulation and coding selection, dynamic resource allocation and load balancing, multi-cell scheduling and mobility management optimizations~\cite{kundu_airan_2025,alzailaa_ranuses_2025}.

However, as 6G \gls{ran} scales with bandwidth, antenna elements, and number of cells, the compute demands of AI-for-RAN workloads grow commensurately~\cite{lin_airanempowering_2025}. In particular, the processing of AI-Native \gls{phy} is subject to tight latency and power consumption constraints. \gls{phy} tasks typically must be completed within 1~ms. To meet this strict time bound, \gls{phy} processing must be executed on the edge, in the base-stations. Additionally, since the evolution to 6G forecasts an increase in cell-site density, base station power consumption must be kept low, allowing only up to 100W power budget for digital signal processing~\cite{bjornson2017massive,Bertuletti_PUSCH_2025}.

In this regard, \glspl{cpu}, \glspl{fpga}, and \glspl{asic} span a wide flexibility versus efficiency tradeoff range, but none alone can unite rapid programmability, elastic compute scaling, low-latency and low-power execution, and adaptability to new workloads or changing requirements required by AI-Native \gls{ran}~\cite{lin_airanempowering_2025}. A new class of programmable \gls{ai}-specialized baseband edge processors is needed, to support high-throughput \gls{phy} workloads within a power envelope of tens of watts, while at the same time providing sufficient flexibility for real-time reconfiguration and long term sustainability. 

\begin{table}[t!] 
\caption{Many-Core Processors for Software-Defined RAN} 
\label{tab:archs} 
\centering 
\resizebox{\columnwidth}{!}{ 
\begin{tabular}{lcccc} 
\toprule 

& \shortstack[c]{TeraPool\\ \cite{Zhang_TeraPool_2025}} 
& \shortstack[c]{X100 \\ \cite{Quallcom_X100}}
& \shortstack[c]{Octeon10\\ \cite{Marvell_Octeon10}} 
& \shortstack[c]{NVIDIA-A100\\ \cite{nvidia_a100}} \\ 
\midrule 

\shortstack[c]{L1-size\\ \quad} 
& \shortstack[c]{4MiB/\\ 1024PEs} 
& \shortstack[c]{-\\ \quad} 
& \shortstack[c]{64KiB/\\PE} 
& \shortstack[c]{128KiB/\\128PE} \\ 

Node 
& 12nm 
& - 
& 5nm 
& 7nm \\ 

Frequency [GHz] 
& 0.88 
& - 
& 2.5 
& 1.41 \\ 

\shortstack[c]{Perf. [TFLOPS@FP16]} 
& 3.6
& - 
& - 
& 78 \\ 

Power [W] 
& 5.5 
& 35 
& 50 
& 400 \\ 
\bottomrule 
\end{tabular} 
}
\vspace{-15pt}
\end{table}

The currently available many-core compute platforms for \gls{phy} processing in \Cref{tab:archs} have the potential to fill this gap: they offer hundreds of general-purpose programmable processors to process in parallel software-defined AI-for-\gls{ran} workloads, non-AI \gls{ran} workloads, and edge AI services on one device. However, these platforms also present limitations. Datacenter \glspl{gpu} show significant power consumption~\cite{nvidia_a100}, making them unsuitable for edge deployments. Smaller many-core \gls{ran} processors~\cite{Quallcom_X100,Marvell_Octeon10,Zhang_TeraPool_2025} achieve lower performance than datacenter \glspl{gpu}, but meet the power consumption requirements of base-stations. Among them, the TeraPool many-core cluster \cite{Zhang_TeraPool_2025} has the lowest power consumption and the highest number of \glspl{pe} with shared low-latency access to L1 memory. This feature allows parallelizing large-dimensional \gls{phy} workloads on hundreds of \glspl{pe}, without fragmenting them across many small clusters, thereby preventing expensive data movement, as well as distribution and synchronization overheads across the memory hierarchy~\cite{Bertuletti_PUSCH_2025}. Nevertheless, the key problem to tackle for \gls{ai}-native \gls{ran} feasibility, is scaling the performance of such many-core baseband processors to \gls{ml} processing needs, at controlled power consumption and high energy efficiency.

From a technology-centric viewpoint, 3D stacking of memory and logic can help meeting the high performance and efficiency requirements of AI-Native \gls{ran} edge workloads. In the high-performance computing market, 2.5D integration with \gls{hbm} is a common solution to feed data-hungry \glspl{npu}~\cite{kwon_hbm_2023,marvell_hbm_2024,prabhakar_sambanova_2025}. However, this approach is expensive and power-hungry (due to the high cost and power consumption of \gls{hbm} memory stacks), and it does not apply to dense edge deployments, such as those envisioned for AI-\gls{ran}. For this reason, 3D stacked chiplet architectures~\cite{wuu_amd_2022}, where multiple memory and/or compute layers are vertically integrated to increase memory capacity and compute density, are expanding from servers toward embedded and edge computing platforms~\cite{amd_epyc4005}.

In this context, we outline our main contributions:
\begin{enumerate}

\item We analyze \gls{ai}-\gls{phy} models according to computational complexity and memory footprint. From this analysis, we identify a significant performance gap for many-core PHY processors when targeting AI workloads, and propose domain-specific architectural specialization focused on \gls{gemm}-dominated computation.

\item We devise TensorPool, an \gls{ai}-native \gls{ran} processor. We integrate 16 \glspl{te} and 256 \glspl{pe} with 4~MiB of shared L1 memory through a hierarchical interconnect, and increase \gls{FP16} peak-performance by 2.25$\times$ (8.4~TFLOPS@\gls{FP16}), compared to a homogeneous cluster of RISC-V \glspl{pe}. 

\item We design a novel microarchitecture for the \glspl{te}’ memory interface to tolerate the latency of the cluster’s physically distributed memory hierarchy and enable multiple \glspl{te} to operate in parallel at high utilization on large \gls{gemm} workloads. In particular, we integrate burst support into the L1 large cluster interconnect to better serve the increased burst bandwidth injected by \glspl{te}. We demonstrate analytically and with RTL-simulation experiments that the proposed architecture is not bound by the L1-memory bandwidth. With a streamlined parallelization scheme, we achieve up to 89\% parallel \gls{fma} utilization on \gls{gemm} and up to 67\% utilization on domain-specific tasks, including data-transfers overheads and concurrent \gls{pe} operation.

\item We demonstrate that a placed\&routed instance of TensorPool in TSMC's 7nm FinFET (N7) node achieves 57.53~GFLOPS@\gls{FP16}/W/mm\textsuperscript{2} on tensor operations (9.1$\times$ higher than the homogeneous cluster baseline, thanks to the increase in compute-units number and utilization). 

\item We show that with 3D-integration and routing in the vertical direction, the area of the routing channels required to connect multiple high-bandwidth \glspl{te} to memory banks can be reduced by 67$\%$, leading to superlinear (2.32$\times$) footprint decrease and opening a promising new avenue for scaling performance and efficiency.
\end{enumerate}

This work demonstrates for the first time that it is possible to meet the throughput and power consumption requirements of upcoming AI-Native \gls{ran} \glspl{phy} with a flexible, heterogeneous shared L1 large-scale cluster with tensor acceleration. Furthermore, we show for the first time that  logic-on-logic 3D integration provides  additional headroom for scaling up the performance and efficiency of shared L1 many-cores.

In \Cref{sec:models} we analyze the compute requirements of AI-\gls{phy}; in \Cref{sec:architecture} we present TensorPool and our novel \glspl{te} to memory high-bandwidth interconnect; in \Cref{sec:memory_balance} we present the analysis of memory balance for the \glspl{te} to L1 connection; in \Cref{sec:performance} we evaluate the utilization of TensorPool compute elements on \gls{gemm}, \gls{phy}, and mixed \gls{te} and \gls{pe} workloads; in \Cref{sec:2D_design} and \Cref{sec:3D_design} we describe \gls{ppa} of respectively our 2D and 3D implementations; in \Cref{sec:soa} we compare the results obtained on TensorPool to \gls{gpu} centric or tensor-based architectures for \gls{ai}-native \gls{phy}; in \Cref{sec:conclusion} we summarize our results and draw conclusions.

\begin{figure}[ht!]
    \centering
    \includegraphics[width=\linewidth]{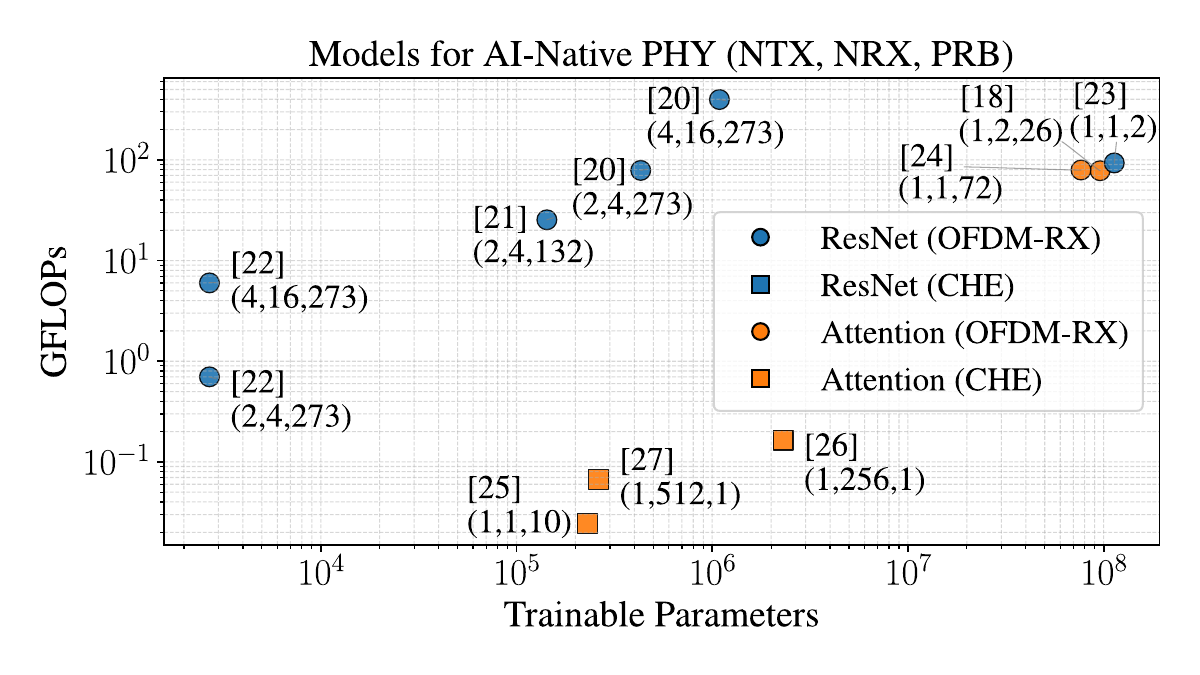}
    \caption{Collection of models for AI-Native PHY.}
    \label{fig:models}
\end{figure}

\section{Models for AI-Native RAN}
\label{sec:models}

In this section, we classify existing \gls{nn}-based models for AI-Native \gls{ran}, to specify the required peak performance and motivate the target workload.

\Cref{fig:models} reports AI-Native \gls{phy} models~\cite{honkala2021deeprx, korpi2021deeprx, cammerer2023neural,wiesmayr_rtnrx_2025,abdollahpour_mdx_2025,wu_aider_2024,wu_darnet_2026,liu_cevit_2023,li_matcenet_2024,wei_xlcenet_2022}, classified according to their architecture, memory footprint (in terms of trainable parameters), number of operations, and target task. We also report the \gls{mimo} size addressed by the model, in terms of number of transmitters, receivers, and \glspl{prb}. All models demonstrated better quality-of-service metrics compared to classical signal processing algorithms. 

Based on their target task, the models can be divided into two groups: models that implement the entire \gls{ofdma} uplink receiver chain~\cite{honkala2021deeprx,korpi2021deeprx,cammerer2023neural,wiesmayr_rtnrx_2025,abdollahpour_mdx_2025,wu_aider_2024,wu_darnet_2026}, and models that implement a single uplink function~\cite{liu_cevit_2023,li_matcenet_2024,wei_xlcenet_2022}. In particular, we focus on models for \gls{che}: the estimation of the transmission channel represents a problem difficult to describe under classical assumptions, especially for large-dimensional channels with moving users, which is a key use-case for AI in \gls{phy} tackled by \cite{li_matcenet_2024,wei_xlcenet_2022}. 

\gls{che} models are not typically used on the full transmission bandwidth, but get dynamically assigned to users requiring a better quality of service in the current transmission slot. For this reason, they were trained on a lower number of \glspl{prb}, compared to full \gls{ofdma} receivers. Nevertheless, we observe that normalizing by the number of \glspl{prb} the number of operations required by \gls{che} models, we obtain a computational complexity comparable to the least expensive full \gls{ofdma} receivers~\cite{abdollahpour_mdx_2025,wiesmayr_rtnrx_2025}. This suggests that a model encompassing the entire uplink can achieve lower computational complexity and memory footprint compared to models trained on a focus task. A flexible processor for AI-Native \gls{ran} should support both approaches. Therefore, we target a computing platform that can jointly perform classical wireless signal processing and \gls{nn} workloads. 

Models on the upper end of computational complexity~\cite{honkala2021deeprx, korpi2021deeprx,cammerer2023neural,wu_aider_2024,wu_darnet_2026} were designed to run on centralized servers or in the cloud~\cite{kelkar_aerial_2021,cohenarazi_aerial_2025}. We decide instead to tune the peak performance of our architecture to the state-of-the-art for real-time edge deployment~\cite{abdollahpour_mdx_2025}. Our real-time reference constraint is the standard 1~ms latency of a \gls{tti}, which contains 14 time-symbols each carrying up to 3276 frequency-orthogonal \glspl{re}, divided in 273 \glspl{prb}. Based on the most demanding use-case of \cite{abdollahpour_mdx_2025} ($N_{TX}=4$, $N_{RX}=16$, $N_{PRB}=273$, 14 time-symbols), we conclude that a many-core for AI-\gls{ran} processing should support at least 6~TFLOPS of AI computations, 1.67$\times$ larger than what is offered by the state-of-the-art TeraPool processor.

In terms of memory footprint, to minimize data movement across the memory hierarchy, the many-core's L1 should contain the \gls{tti} input samples and the model parameters. Excluding the models for cloud deployment, and assuming \gls{FP16} arithmetic precisions, all other approaches fit in 4~MiB.

Finally, we observe that the proposed survey contains both ResNet-inspired convolutional models~\cite{honkala2021deeprx, korpi2021deeprx,cammerer2023neural,wiesmayr_rtnrx_2025,abdollahpour_mdx_2025} and attention-based models~\cite{wu_aider_2024,wu_darnet_2026}, both heavily dominated by \gls{gemm} computations. Therefore, we can obtain the required improvements in computational complexity, while keeping low power consumption, by domain-specialization of our architecture focused on \gls{gemm} operations.

\section{Architecture}
\label{sec:architecture}

Based on the analysis of state-of-the-art models for AI-Native \gls{phy}, we design TensorPool, a domain-specific many-core heterogeneous architecture combining general-purpose \glspl{pe} and \glspl{te}. TensorPool architecture achieves $>$8~TFLOPS@\gls{FP16} peak performance, by integrating domain-specific \gls{gemm} accelerators with 4~MiB of shared multi-banked L1 scratchpad.

\subsection{TensorPool's Interconnect}

The \textit{Pool}~\cite{Zhang_TeraPool_2025} is a modular design, where hundreds of \gls{pe} are tightly-coupled to a shared multi-banked scratchpad memory and fetch the data in parallel through a low-latency (1-9 cycles), hierarchical, \gls{xbar} based interconnect.

\begin{figure}[ht!]
    \centering
    \includegraphics[width=\linewidth]{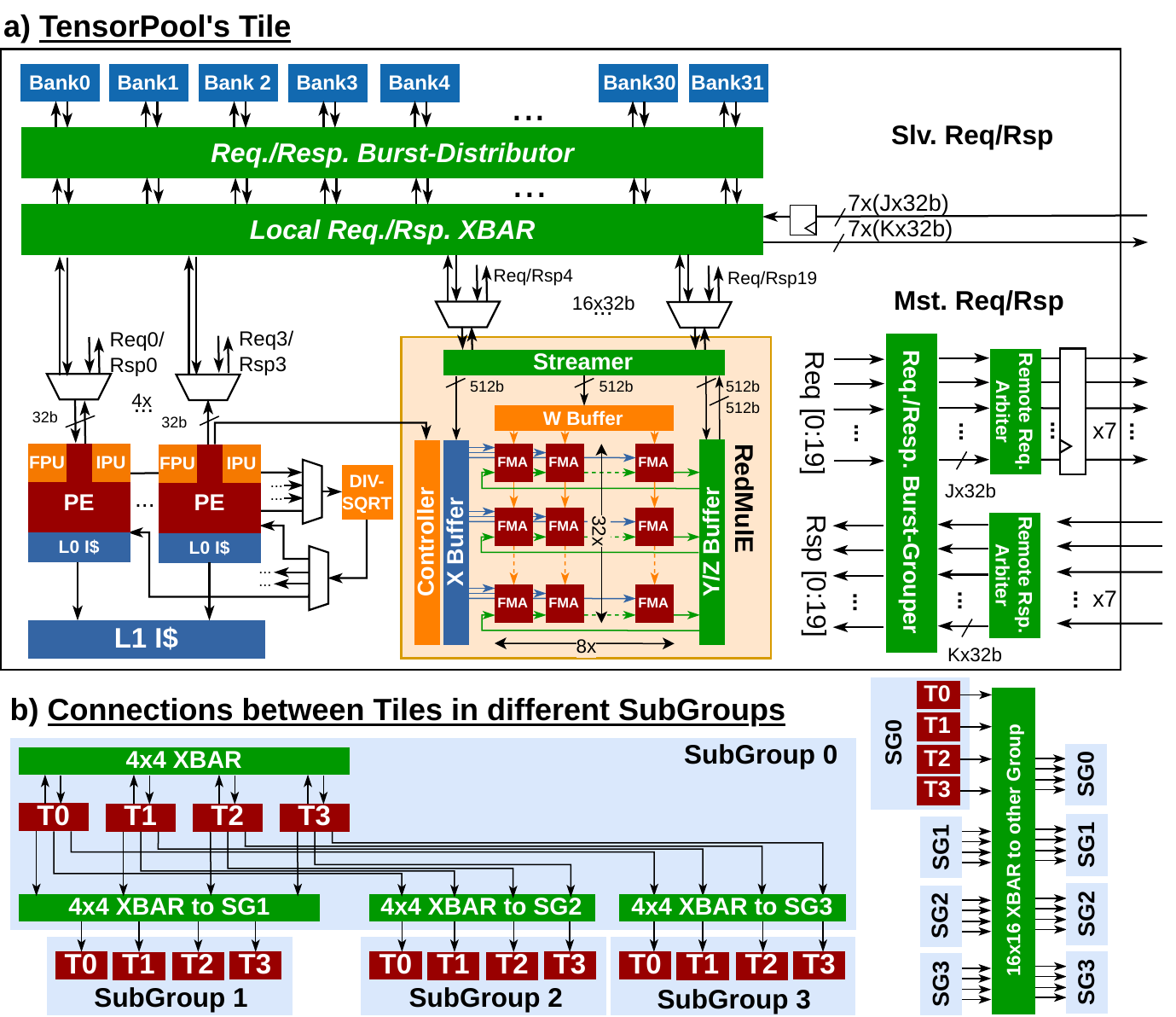}
    \caption{(a) TensorPool Tile, (b) Connections between Tiles in TensorPool via hierarchical crossbars.}
    \label{fig:architecture}
\end{figure}

The Pool is assembled bottom up from \textit{Tiles}. A Tile in \Cref{fig:architecture}-a, contains 4 \textit{RV32IMAZfinx} \glspl{pe}, augmented with baseband domain-specific ISA-extensions~\cite{Bertuletti_PUSCH_2025}.
Each \gls{pe} is capable of two \gls{FP16}-MACs/cycle in parallel on their 32-bit \gls{fpu}, a shared Div-Sqrt \gls{fpu}, 32$\times$2KiB memory banks, and 4~KiB shared L1-I\$. Tile banks are accessed in one cycle through a local \gls{xbar}. The L1 in other Tiles is accessed through a remote-request arbiter. TensorPool contains 64 Tiles, divided into three hierarchies and connected via low-latency \glspl{xbar}, as in \Cref{fig:architecture}-b, to keep routing complexity low and ensure physical feasibility of the design: 4 Tiles form a \textit{SubGroup}, and 4 SubGroups make a \textit{Group}. To ease timing-closure, spill-registers are added at the Tile, Group, and SubGroup boundaries. The \glspl{pe} access latency to the distributed scratchpad banks is 3 cycles within the SubGroup, 5 within the Group, and 9 to other Groups.

The number of Tiles and \glspl{pe} in TensorPool is selected based on the results obtained by \cite{Zhang_TeraPool_2025,shen_bursts_2025}, where the routing congestion sets a $32\times32$ upper limit to the \glspl{xbar} size. In TensorPool, we reduce the size of the largest \gls{xbar} to $16\times 16$ and leave the headroom to integrate \glspl{te} with wide memory access ports.

\begin{figure}[ht!]
    \centering
    \includegraphics[width=0.65\linewidth]{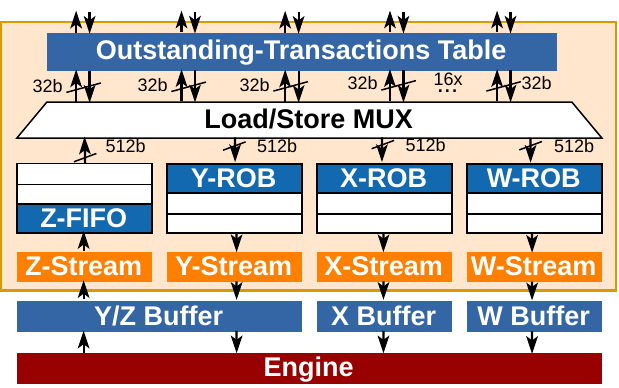}
    \caption{RedMulE outstanding streamer and transactions table.}
    \label{fig:streamer}
\end{figure}
\begin{figure}[ht!]
    \centering
    \includegraphics[width=0.8\linewidth]{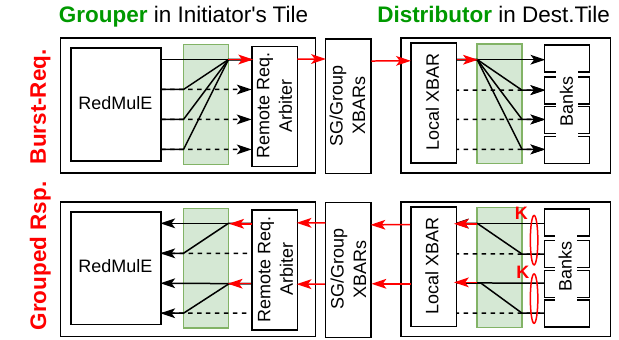}
    \caption{Bursting of the reads and re-distribution to target banks. Scheme for response grouping on the same handshake.}
    \label{fig:bursts}
\end{figure}

\subsection{Latency Tolerant TE Integration}

To boost performance on \gls{gemm} ($Z=Y+X*W$), a Tile per SubGroup contains a \gls{te} with 256 \glspl{fma}. The \glspl{fma} add up to the \glspl{pe} compute capabilities, delivering peak 288 \gls{FP16}-MACs/cycle combined throughput, 2.25$\times$ larger than in TeraPool's 64-\glspl{pe} SubGroup. 

In the \gls{te}, \glspl{fma} are arranged in $R = 32$ rows by $C = 8$ columns. Three data buffers (Y\&Z share the same) fed by a streamer provide data to the compute elements. The streamer loads and stores data through a $C\times(P+1)\times\text{16-bit}=\text{512-bit}$ wide memory port. Each \gls{te}-row computes a dot-product between an X-row and a W-column, accumulating in a Y-element. On each \gls{te} column, the same X value is kept stationary and is multiplied by different columns of W. The \glspl{fma} have $P = 3$ pipeline stages. At each cycle they pass over the partial dot-product to the next pipeline stage. To keep \glspl{fma} always busy, the \gls{te} must load from W $C\times(P+1)$ elements every four cycles, while it refills the X and Y buffers. The \gls{te}'s controller exposes a set of configuration registers, programmed in software by one of the Tile's \glspl{pe}. After programming the \gls{te}, the \gls{pe} is free to execute independent code upon receiving an interrupt request by the \gls{te} on workload completion.

A key challenge in TensorPool is to connect 16 \glspl{te} to all the distributed scratchpad memory banks through the hierarchical multi-cycle interconnect, while hiding memory latency and achieving full \gls{fma}-utilization. In \cite{tortorella_redmule_2023} and \cite{yi_opengemm_2025}, the latency of tightly coupled memories is hidden by data buffers: a streamer is allowed continuous pre-fetching of data until the buffer reaches its capacity. However, the streamer assumes in-order responses, which is unfortunately not what happens in a system with non-uniform memory latencies. 

To overcome this mismatch, which would lead to significant TE stalls and underutilization, we designed the latency-tolerant streamer in \Cref{fig:streamer}. It includes a 16-entry \gls{rob} on each stream, X, W, Y. The \gls{rob} keeps track of in-flight reads and allows to pipeline multiple outstanding transactions in the multi-cycle interconnect. A transactions table is also necessary to collect 32-bit bank responses, and commit 512-bit wide responses to the \gls{te}. Finally, a 32-entry FIFO on the Z stream allows us to empty the Y/Z buffers upon dot-product completion and immediately pre-load new Ys.

 The Tile's arbiter can only retire up to 7 transactions per cycle, four addressing the Tiles of SubGroups within a Group and 3 addressing Tiles in remote Groups. This represents a bottleneck for the wide \glspl{te} requests leaving the Tile, which get narrowed to the arbiter's bandwidth and are dispatched in multiple cycles. This creates backpressure on the streamer and blocks new requests. 
 
 To avoid serialization of wide requests at the Tile's arbiter, we take advantage of the burst-based access patterns of \glspl{te} to memory, and augment the interconnect with lightweight burst support~\cite{shen_bursts_2025}. As shown in \Cref{fig:bursts}, a \textit{Burst-Grouper} in the initiator's Tile intercepts a 512-bit wide read-request, transforming it into a \textit{burst}: only the address to the first 32-bit element in the 512-bit request is sent to the arbiter and crosses the hierarchical interconnects. Subsequent wide requests are let through the arbiter in the next cycle, restoring a peak throughput of 7 wide read transactions per cycle. A \textit{Burst-Distributor} in the destination Tile identifies burst-requests and generates 16 narrow requests to target banks.

Serialization of responses at the local crossbar of the destination Tile can also be partially avoided by increasing the interconnect bandwidth. We achieve that while avoiding the introduction of additional ports in the crossbars by widening the data field only of our custom protocol. In fact, we group \textit{K} responses under the same valid/ready handshake, limiting the impact of additional read bandwidth on the routing complexity. \Cref{fig:bursts} is an example of the adopted strategy for $K=2$. A similar approach is also adopted for wide write-requests: the width of the request data field is multiplied by a factor $J=2$.

\subsection{L2 Memory Interconnect}

The system is completed by a top-level AXI-\gls{xbar}, accessed in parallel by all Tiles through the hierarchical AXI-interconnect described in~\cite{Zhang_TeraPool_2025}, and delivering up to 512-bit/cycle read\&write bandwidth to each SubGroup. The \gls{xbar} connects TensorPool to L2 and allows any \gls{pe} to program a centralized \gls{dma} engine through a register-based interface. The \gls{dma} sends data requests to L2 and redistributes responses through the hierarchical AXI-interconnect to the memory banks in all Tiles.

\section{Memory Balances}
\label{sec:memory_balance}

In this section, we demonstrate that thanks to the high-bandwidth L2 memory connection, the \glspl{te} are not bound by L2 memory transfers. Moreover, we demonstrate that thanks to the multiple-outstanding burst-transaction feature and the enlarged response bandwidth of shared L1 memory interconnects, the single-\gls{te} is not bound by L1 memory latency. These theoretical results are backed by experimental validation, where we demonstrate near-ideal utilization of the single-\gls{te}'s \glspl{fma}.
\subsection{Memory Balances According to Kung's Principle}
Kung's principle~\cite{kung_1986} argues that to ensure full utilization of compute resources, their speed should not outpace memory bandwidth. In other words, for a fixed workload, the time required to transfer data from higher memory hierarchies should never exceed the time required to process these data.
\subsubsection{L2-balance}
It is easy to demonstrate that TensorPool is not bound by L2 memory transfers according to Kung's principle. The Pool's peak performance is $\pi_{TEs} = 16 \times 256~\text{MACs/cycle}$. Its L2 read\&write bandwidth is $\beta_{L2} = 1024~\text{B/cycle}$. Taking into account a squared ($n \times n \times n$) \gls{FP16} \gls{gemm}, executed double-buffering computation and memory transfers, we compute the total number of arithmetic operations $Wk$, and the total number of bytes in-flight $Q_m$, then we write Kung's inequality:

\begin{equation}
\begin{aligned}
&Wk = n^3~MACs\\
&Q_m = 2B \times (n^2 + n^2 + 2 \times n^2) = 8n^2~B\\
&\left( T_{compute} = \frac{Wk}{\pi_{TEs}} = \frac{n^3~MACs}{8192~MACs/cycle} \right) \geq\\ 
&\left( T_{transfer} = \frac{Q_m}{\beta_{L2}} = \frac{8n^2~B}{1024~B / cycle} \right)\\
\label{eq:gemm_balance}
\end{aligned}
\end{equation}

The double-buffering assumption requires half of the memory to be transferred: $Q_m = 8n^2 B = 2MiB \rightarrow n=512$. For this problem size, Kung's inequality holds.

\subsubsection{L1-balance within a Tile}
We also want to demonstrate that a single \gls{te} is not bound by the tightly coupled local interconnect to the L1-memory within a Tile. To demonstrate this result, it is enough to demonstrate that the inner loop of RedMulE's computation is balanced. During the inner loop, RedMulE computes a tile of the output matrix with $C \times R \times (P+1)$ elements. It fetches $R$ rows from X and $C \times (P+1)$ columns from W. Assuming $n$ elements per row and column, this corresponds to a $R \times n \times C(P+1)$ \gls{gemm}, which according to \Cref{eq:gemm_balance} yields:
\begin{equation}
\begin{aligned}
Wk &= \left[ R \times n \times C(P+1) \right]~MACs = 1024n~MACs \\
Q_m &= 2B \times \left[ nR + nC(P+1) + 2 \times RC(P+1) \right]\\ 
&= (128n + 2048)~B \\
\end{aligned}
\end{equation}

To simplify the calculations, we can drop the $Q_m$ term that does not depend on $n$. This means that the following considerations hold strong for a large enough problem over the dimension of the dot-products executed by the accelerator. The peak-performance offered by a single \gls{te} are $\pi_{TE} = 256~\text{MACs/cycle}$. The memory bandwidth between the \gls{te} and the local memory in its Tile is $\beta_{loc.} = 512~\text{bit/cycle} = 64~\text{B/cycle}$. If the input matrices are allocated within a Tile, the connection is not memory-bound:
\begin{equation}
\begin{aligned}
&\frac{Wk}{\pi_{TE}} \geq \frac{Q_m}{\beta_{loc.}} \rightarrow \frac{\pi_{TE}}{\beta_{loc.}} \leq \frac{Wk}{Q_m} \rightarrow \frac{\pi_{TE}}{\beta_{loc.}} \leq 8~MACs/B\\
\end{aligned}
\end{equation}

\subsubsection{L1-balance outside of a Tile}
We finally demonstrate that a single \gls{te} is not bound by the hierarchical interconnect to the distributed L1-memory in other Tiles. To further simplify the calculation, we assume that the \glspl{te}' 512-bit wide accesses all have a random starting address. The assumption works well when the matrices are large enough to be well distributed across all memory banks and when the Xs and Ws do not overlap, which can be easily avoided using the workload mapping strategies explained later in \Cref{sec:performance}. We indicate with $N_{B}=2048$ and $N_{B/T}=32$ the total banks of TensorPool and the banks in a Tile. We first compute the probability of local and remote access ($p_{loc.}$ and $p_{rem.}$). We then combine the bandwidth to banks within a Tile and the bandwidth to remote locations (through the Tile's arbiter), to compute the total bandwidth for a \gls{te} ($\beta$) and the complete (local and remote access) Kung's inequality.
\begin{equation}
\begin{aligned}
&p_{loc.} = \frac{N_{B/T}}{N_{B}}, \quad p_{rem.} = (1 - p_{loc.})\\
&\beta = p_{loc.} \times \beta_{loc.} + p_{rem.} \times \beta_{rem.} \\
& \frac{\pi_{TE}}{\beta} \leq \frac{Wk}{Q_m} \rightarrow \frac{\pi_{TE}}{\beta} \leq 8~MACs/B
\end{aligned}
\end{equation}

The local bandwidth $\beta_{loc.}$ is known. To compute the remote bandwidth $\beta_{rem.}$ we need to consider the bandwidth of each remote port: $\beta_{port} = K\times4~\text{B/cycle}$. To fulfill the inequality, at least two ports must be active together in every cycle. By selecting $K=4$, thanks to the outstanding transaction support, \glspl{te} can send one request per cycle and keep a remote port busy for $64B/\beta_{port}=4~\text{cycles}$. We indicate with $N_{B/G}=512$ the banks in a Group, with $N_{G}=4$ the number of Groups, and with $N_{SG/G}=4$ the SubGroups in a Group. The probability that in four cycles all random requests target the same remote port is:
\begin{equation}
\begin{aligned}
&p^{*} = \frac{3N_{B/G}}{N_B}\left(\frac{1}{N_G}\right)^3 + \frac{N_{B/G}}{N_B}\left(\frac{1}{N_G}\frac{1}{N_{SG/G}}\right)^3 = 0.012
\end{aligned}
\label{eq:p_star}
\end{equation}

The first term of the sum in \Cref{eq:p_star} represents random requests to the three Group ports of a Tile. It is weighted by the probability that in four cycles, three requests after the first one all target the same port. The second term represents the accesses to the four SubGroup ports of a Tile. It is weighted by the probability that the three following independent random requests all target the same SubGroup. The probability that at least two ports are active per cycle is $1-p^{*}=0.988$. The bandwidth for remote accesses $\beta_{rem.}$ will always be higher than the value $\beta^*$ calculated for two simultaneous active ports at most. This demonstrates that selecting $K=4$, Kung's inequality is fulfilled across both local and remote memory accesses, and the architecture is not memory-bound:
\begin{equation}
\begin{aligned}
&\beta_{rem.} > p^{*} \times \beta_{port} + (1-p^{*})\times2\beta_{port} = \beta^{*} \\
&\beta > p_{loc.} \times \beta_{loc.} + p_{rem.} \times \beta^{*} \\
&\frac{\pi_{TE}}{\beta} < \frac{\pi_{TE}}{p_{loc.} \times \beta_{loc.} + p_{rem.} \times \beta^{*}} < 8 MACs/B
\end{aligned}
\end{equation}

\begin{figure}[t!]
  \centering
  \includegraphics[width=\columnwidth]{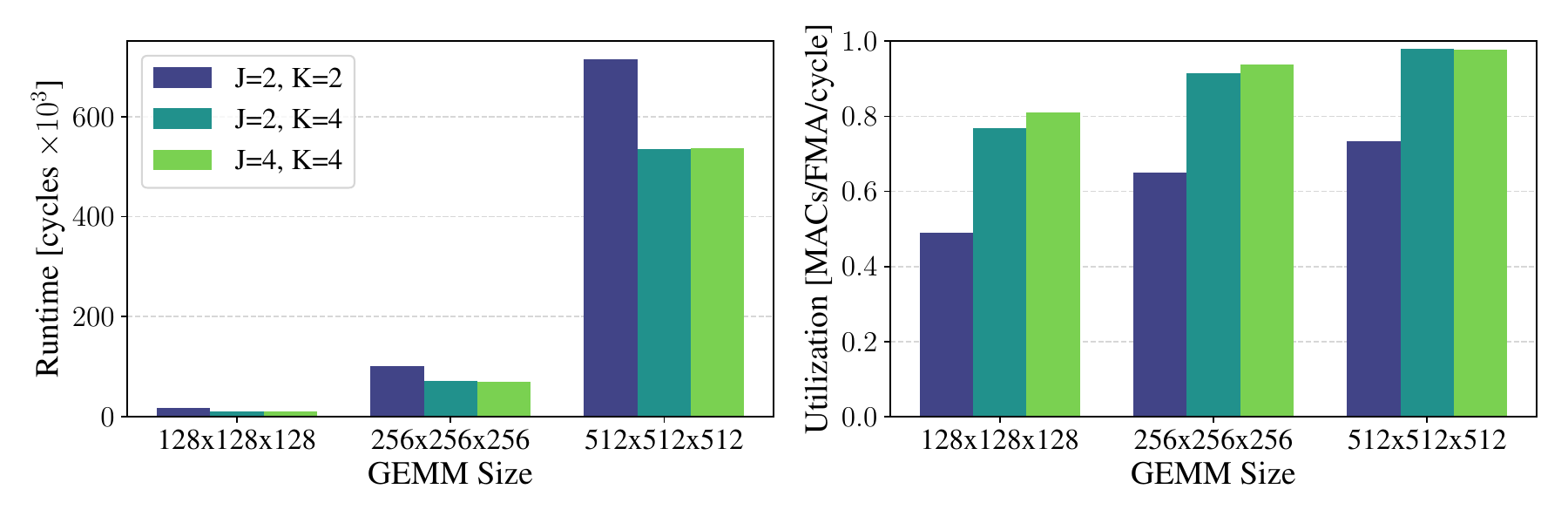}
  \caption{Single-\gls{te} \gls{gemm} performance as a function of problem size and of interconnect bandwidth.}
  \label{fig:single_gemm}
\end{figure}

\subsection{Single-TE Performance on Interconnect Bandwidth Scaling}

In \Cref{fig:single_gemm} we benchmark the runtime and \gls{fma} utilization of a single-\gls{te} on different sizes of \gls{gemm}. We also vary the grouping request and response factors J and K, progressively increasing the tightly-coupled scratchpad write and read bandwidth. Performance is measured with cycle-accurate \gls{rtl} simulation in QuestaSim 2022.3.

As shown in \cite{tortorella_redmule_2023}, the \gls{te} utilization increases with the problem size: the time required to fill the accelerator's pipeline at each inner loop iteration is amortized on larger problem sizes. We also observe that increasing the interconnect bandwidth reduces the congestion on the Tile's remote arbiters, leading to higher utilization. The utilization peaks at 98\% for a large problem size, $J=2$ and $K=4$. Further increasing the remote bandwidth would only exacerbate the routing congestion at the Pool's hierarchies' boundaries.

These experiments on the single-\gls{te} validate the analytic results obtained by computing the L1-memory balance: TensorPool achieves near-to-ideal peak performance for $K=4$.

\begin{figure}[t!]
    \centering
    \resizebox{\linewidth}{!}{
        \includegraphics[width=\linewidth]{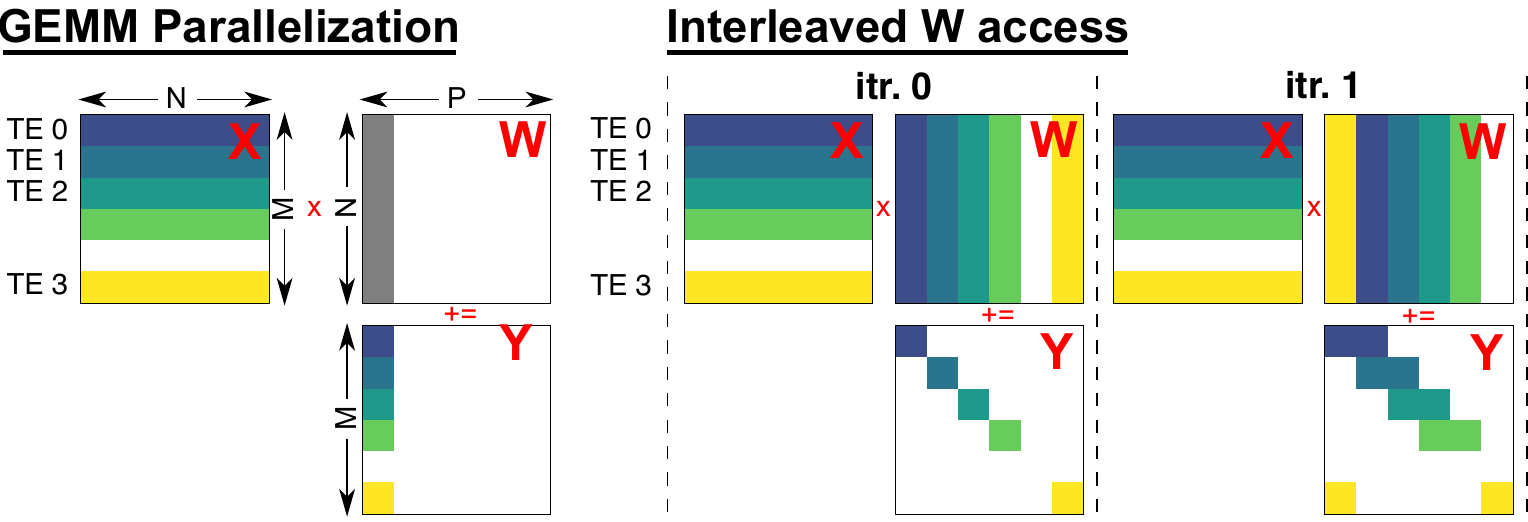}
    }
    \caption{GEMM Parallelization across \glspl{te} and interleaved access to W columns, aiming at reduced bank-contentions.}
    \label{fig:gemm}
\end{figure}
\begin{figure}[t!]
  \centering
  \includegraphics[width=\columnwidth]{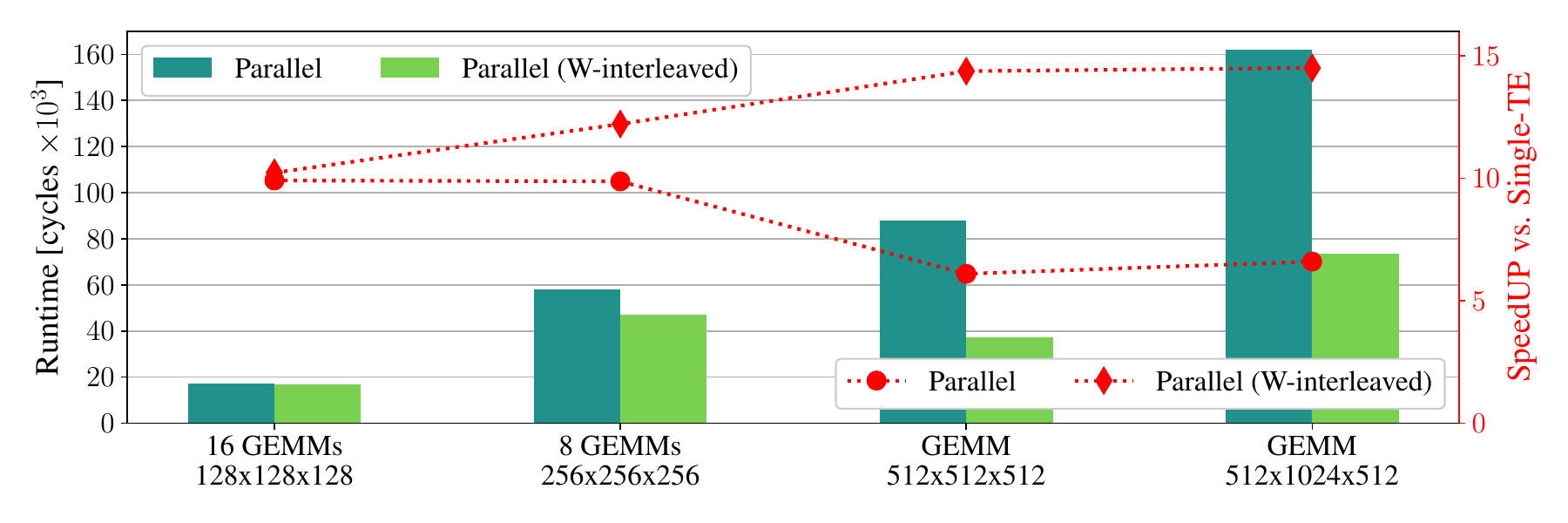}
  \includegraphics[width=\columnwidth]{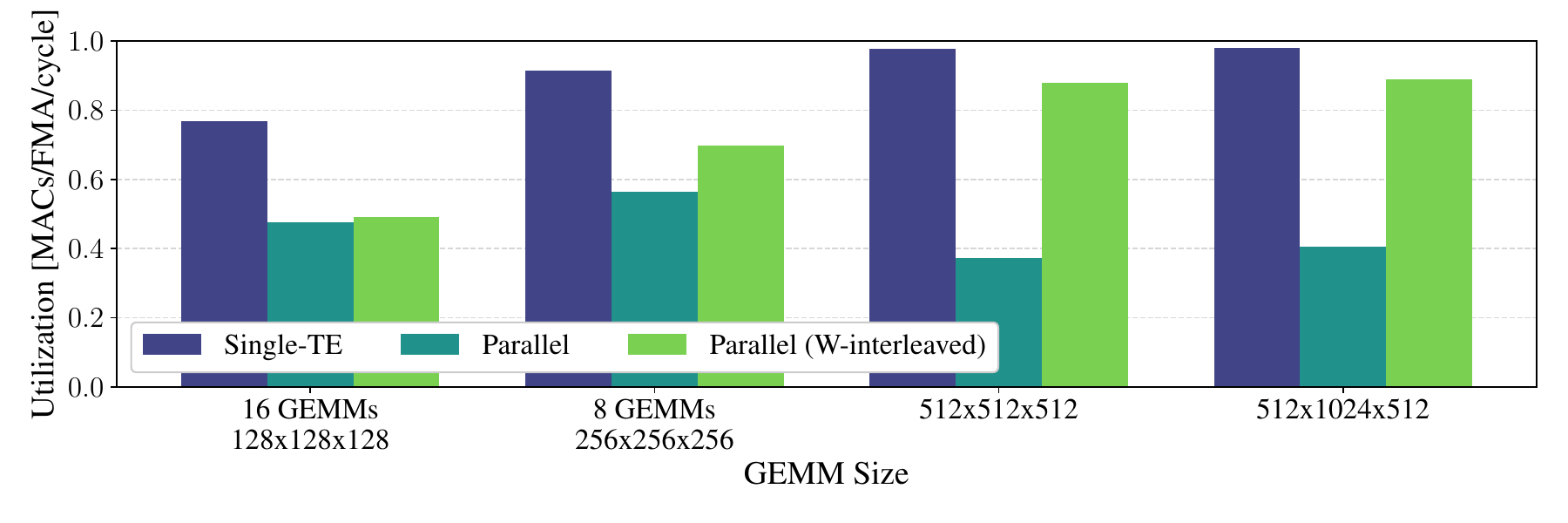}
  \vspace{-10pt}
  \caption{Runtime and utilization of parallel \gls{gemm} on 16 \glspl{te}.}
  \label{fig:gemm_perf}
\end{figure}
\begin{figure}[t!]
  \centering
  \includegraphics[width=\columnwidth]{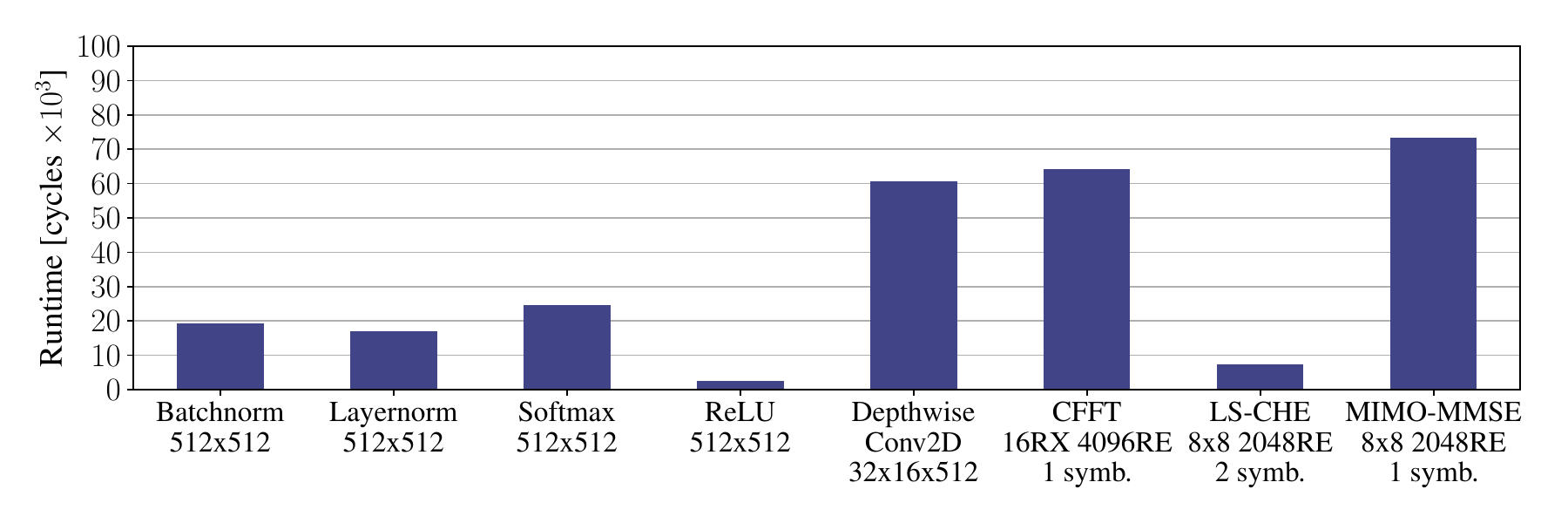}
  \includegraphics[width=\columnwidth]{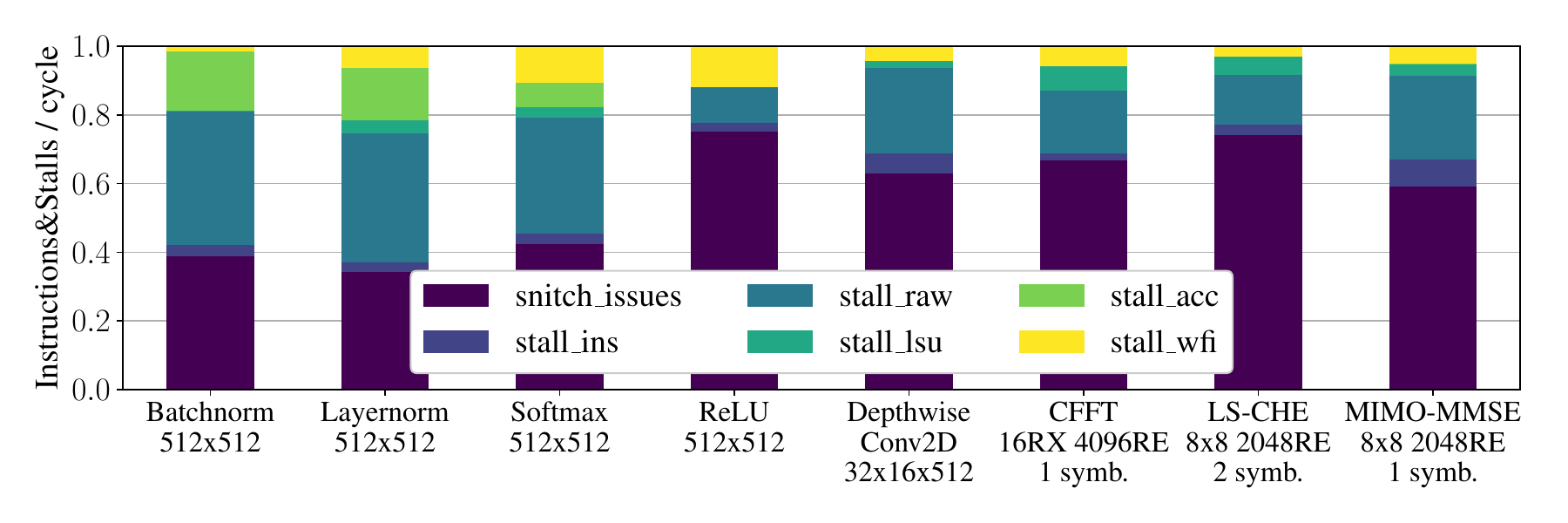}
  \caption{Runtime, instructions\&stalls per cycle breakdown for various parallel AI-Native \gls{phy} and classical signal-processing kernels optimized for the \glspl{pe} of TensorPool.}
  \label{fig:kernels_perf}
\end{figure}

\section{Workload Mapping \& Cycle-Count Performance}
\label{sec:performance}

In this section, we discuss how one can map large-dimensional AI-Native \gls{phy} and classical wireless processing workloads to TensorPool, fully exploiting the peak compute capabilities of the \glspl{te} and \glspl{pe} parallel workforce.

\subsection{Performance of \gls{gemm} on Parallel TEs}

In TensorPool, the execution of \glsentryshort{gemm} is flexible and benefits from multiple degrees of parallelism. Each \gls{te} can independently execute a \glsentryshort{gemm}, alternatively, a large \glsentryshort{gemm} residing entirely in the cluster L1 can be parallelized across 16 \glspl{te}, as in \Cref{fig:gemm} on the left (each \gls{te} computes dot-products for a subset of rows in Z, while fetching from corresponding rows in Y and X, and the entire W).

To avoid bank-access contentions on parallel access to W, we develop the interleaved access scheme in \Cref{fig:gemm} on the right. Each \gls{te} can start computing a tile of the output matrix from a different column of W and then loop back to the first column at the end of the partial computation. The starting address on the W columns is written by the core in the \glspl{te} configuration registers along with the other parameters. Support for the loop back feature is added to the state machine of the controller that governs the \glspl{te} operation.

\Cref{fig:gemm_perf} shows the runtime and utilization of multiple parallel independent \glsentryshortpl{gemm} and large \glsentryshort{gemm} workloads distributed between 16 \glspl{te}. We report up to 14.5$\times$ speedup compared to execution on a single RedMulE. We show that thanks to a low-latency interconnect, the multiple outstanding transaction support of the new RedMulE streamer, burst support, and an efficient parallelization scheme, TensorPool sustains parallel tensor computations concurrently on all the \glspl{te} with up to 89\% \gls{fma}-utilization, and small synchronization overhead, even in comparison to a single execution (98\%). In particular, interleaving the access to W columns boosts the parallel \gls{fma}-utilization of up to 48\% on large matrices.

\subsection{Performance of AI-Native \gls{phy} and Classical Wireless Signal-Processing Parallel Workloads on PEs}

Thanks to the inherent flexibility of a large shared L1 architecture, the RISC-V general-purpose \glspl{pe} of TensorPool execute independent computations in parallel to \glspl{te}. TensorPool employs a simple fork-join programming model in which each \gls{pe} and \gls{te} is assigned a portion of the input data according to its unique ID (exposed via a lightweight bare-metal API). Processing is performed in parallel on shared L1 data, followed by synchronization.

In \Cref{fig:kernels_perf}, we report the cycle runtime and a breakdown of instructions and stalls over the total runtime for key AI-Native \gls{phy} and classical wireless signal-processing workloads, optimized to run in parallel on \glspl{pe} of TensorPool.

The RISC-V general-purpose \glspl{pe} of TensorPool \circled{1} complete the \gls{te} functionalities, processing auxiliary \gls{nn} operations, such as activations, and \circled{2} offer the flexibility required to execute diverse model-driven wireless signal processing workloads.

Regarding the first point, we observe that parallel Batchnorm, Layernorm, Softmax, and \gls{relu} have smaller runtime than an equal-size \glsentryshort{gemm}, enabling overlapped execution of dense-layers and activations in a \gls{nn}. 

Regarding the second point, we note that many classical signal-processing algorithms, such as \gls{cfft}, \gls{ls} \gls{che}, and \gls{mimo} \gls{mmse} detection cannot benefit from acceleration on the \glspl{te}. However, these functions should still be supported by a programmable base-station platform for AI-Native \gls{ran}. On the one hand, there are \gls{nn} models that only apply to a section of the uplink \gls{phy}. On the other hand, to maintain the computational complexity of the uplink under control, the network should resort to full expensive \gls{ofdma} receivers only in critical scenarios, where the quality of service would be severely compromised based on classical signal-processing alone~\cite{zayene_2025_adaptive_nw}. Notably, \glspl{pe} achieve respectively, 0.77, 0.59, and 0.66 instructions/cycle on parallel \gls{che}, \gls{mimo}-\gls{mmse}, and \gls{cfft}. Assuming 0.9~GHz operation (aligned with previous physical implementations of the \textit{Pool} interconnect~\cite{Zhang_TeraPool_2025}), TensorPool can address \gls{cfft}-based \gls{ofdma} for $N_{TX}=16$, 4096 \glspl{re}, and 14 symbols in 1~ms (corresponding to 273 \glspl{prb}, the highest allowed by 5G specifications). The 8$\times$8 \gls{mimo} \gls{mmse} detection and \gls{che} on 14 symbols with 2048~\glspl{re} each are completed in 0.97~ms, within the 1~ms \gls{tti} constraint.

\begin{figure*}[ht!]
  \centering
  \includegraphics[width=\textwidth]{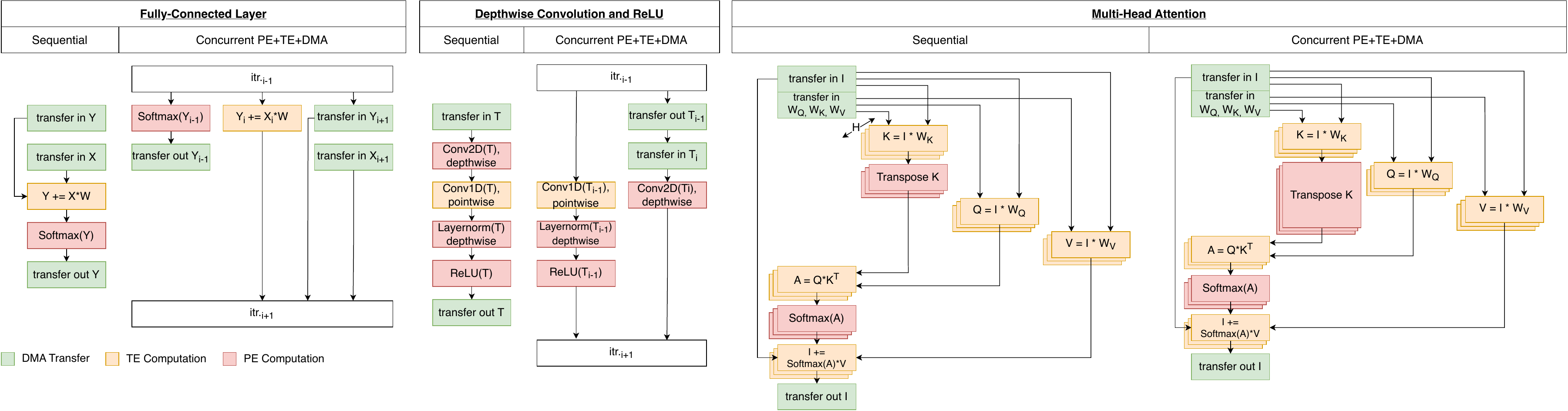}
  \caption{Data-flow diagram for the sequential and a concurrent (\glspl{te}, \glspl{pe}, \gls{dma}) execution of AI-Native \gls{phy} compute blocks.}
  \label{fig:model_diagrams}
\end{figure*}
\begin{figure*}[h!]
  \centering
  \includegraphics[width=0.329\textwidth]{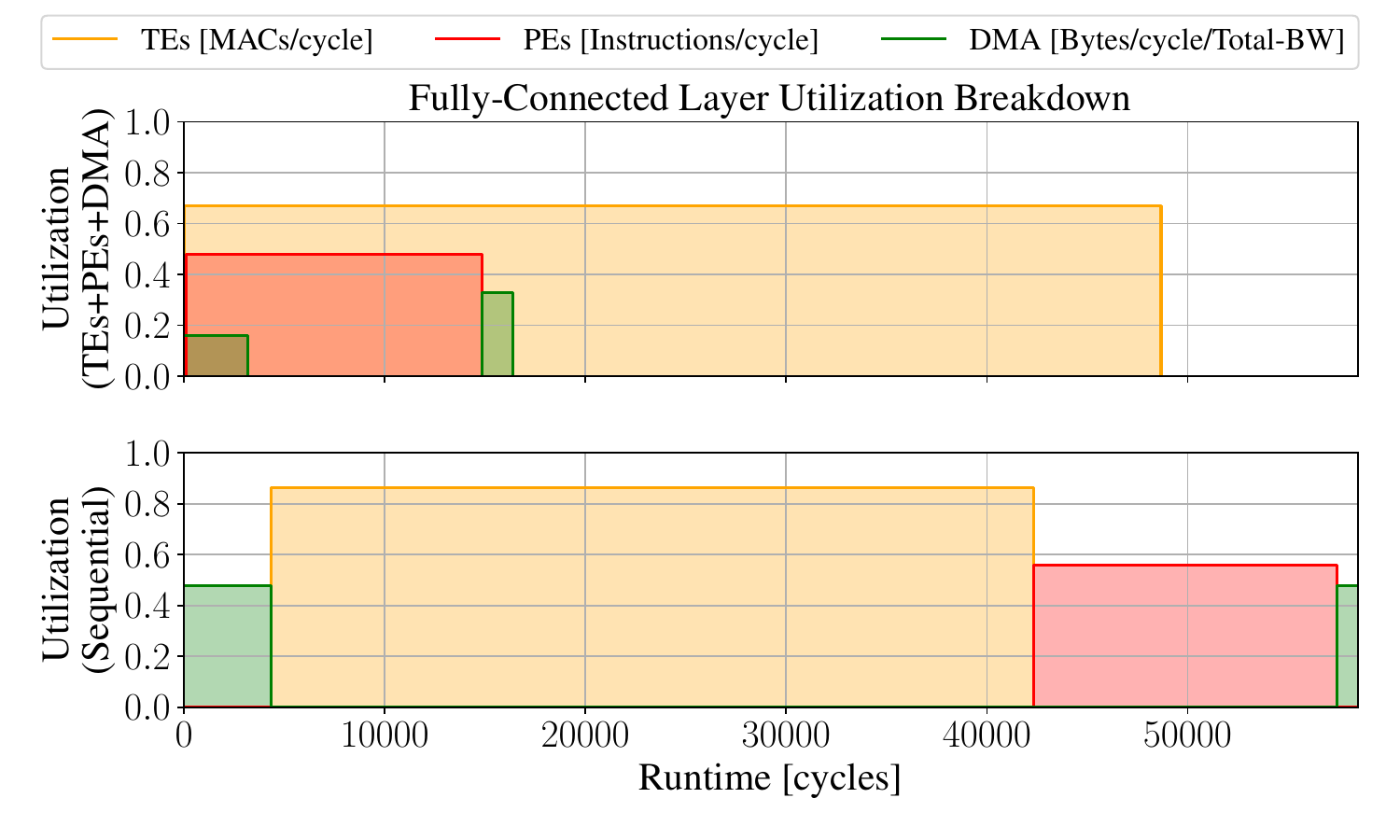}
  \includegraphics[width=0.329\textwidth]{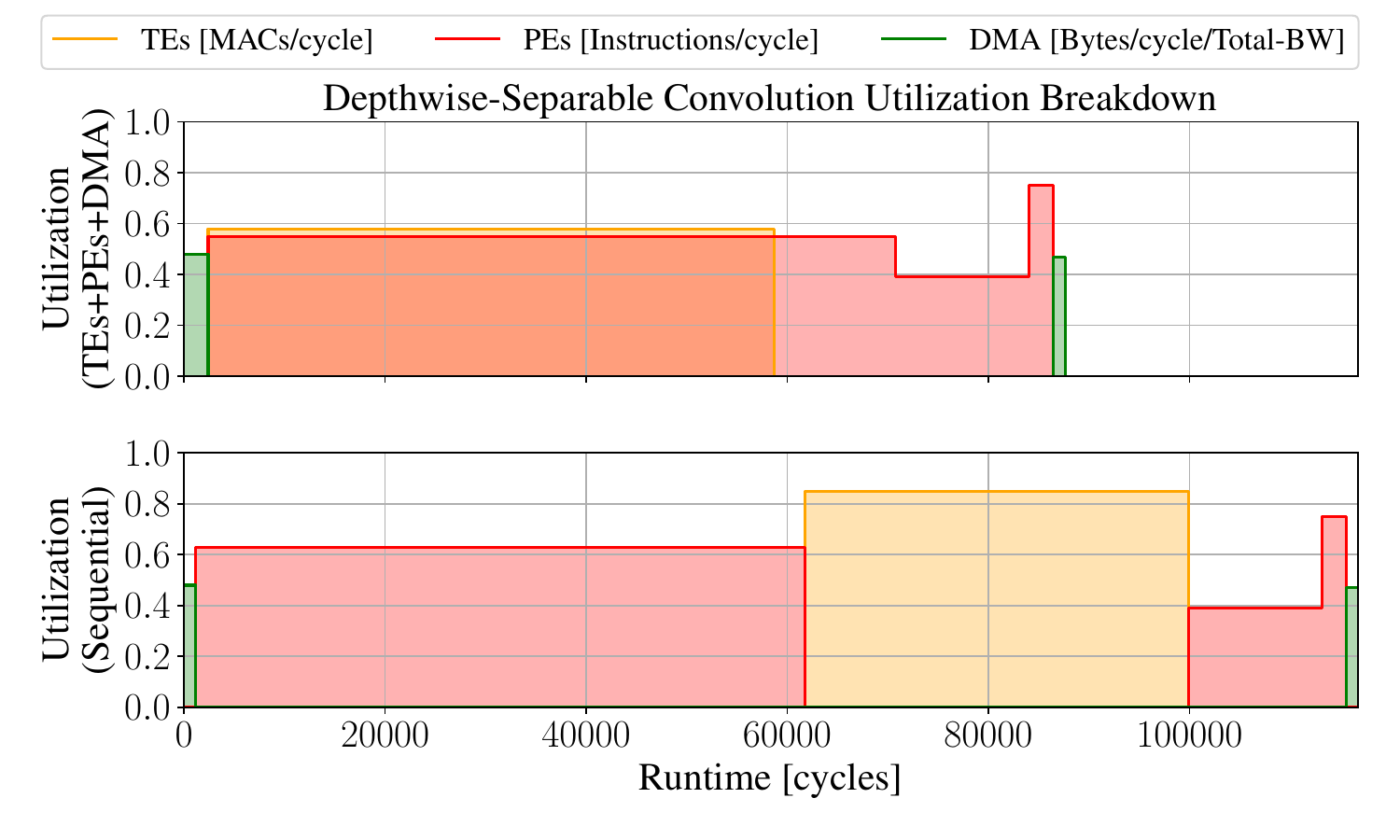}
  \includegraphics[width=0.329\textwidth]{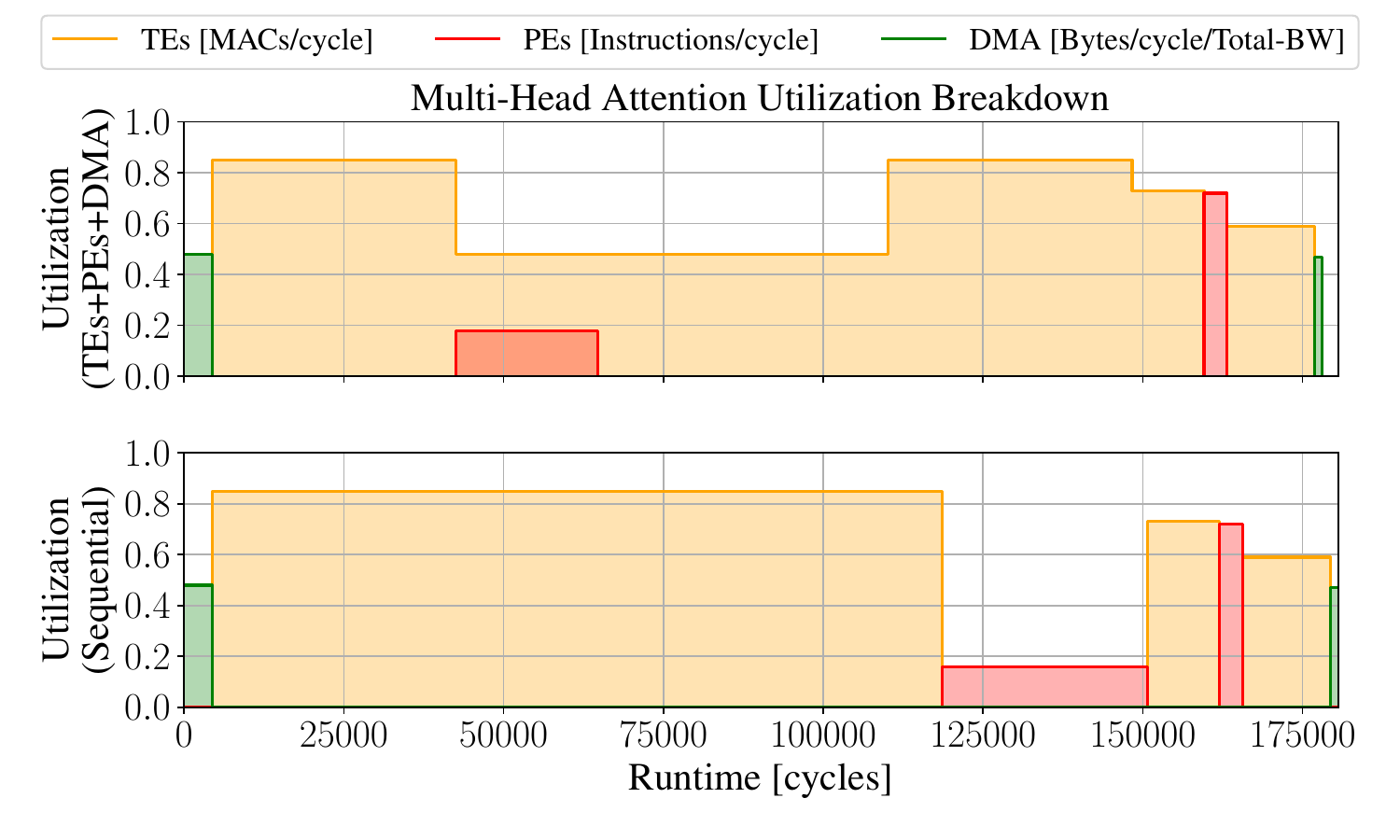}
  \caption{Runtime and utilization of \glspl{te}, \glspl{pe}, \gls{dma} in sequential and concurrent execution of AI-Native \gls{phy} compute blocks.}
  \label{fig:temporal_diagram}
\end{figure*}

\subsection{Performance of Models for AI-Native \gls{phy} Executed in Parallel on TEs\&PEs}

We benchmark the performance of TensorPool in the case of concurrent operation of \glspl{te}, \glspl{pe} and \gls{dma}. We focus on the three use-cases represented in \Cref{fig:model_diagrams}: a \gls{fc} layer with \gls{gemm} and softmax activation, a depthwise-separable convolution with layer normalization and \gls{relu}, a \gls{mha}. These compute blocks are at the core of the models described in \Cref{sec:models}. For each block, we propose a sequential implementation, where we operate \glspl{te}, \glspl{pe} and \gls{dma} one at a time, and a concurrent implementation, where computation and data transfer can overlap in time. The temporal diagram of the computations and transfers divided on \glspl{te}, \glspl{pe}, and \gls{dma} is in \Cref{fig:temporal_diagram}.

The \gls{fc} layer with row-wise softmax activation is used in all models in our survey. For the concurrent operation of \glspl{pe}, \glspl{te}, and \gls{dma}, we propose a double-buffer implementation: while \glspl{te} compute \gls{gemm} on the data of the current iteration, \glspl{pe} execute softmax on the result of the previous \gls{gemm} iteration. The \gls{dma} transfers the input of the next iteration to L1, and it stores the softmax output to L2 after \glspl{te} and \glspl{pe} complete the processing. The runtime results in \Cref{fig:temporal_diagram} correspond to a 512$\times$512 input matrix.

The depthwise-separable convolution with layer normalization and \gls{relu} is the bulk of the ResNet models for \gls{ofdma}~\cite{honkala2021deeprx,korpi2021deeprx,cammerer2023neural,wiesmayr_rtnrx_2025,abdollahpour_mdx_2025,wu_aider_2024,wu_darnet_2026}. It consists of a depthwise 2D-convolution, and a pointwise 1D-convolution. The 1D-convolution is a matrix-matrix multiplication running on \glspl{te}, with accumulation along the input tensor depth. The depthwise 2D-convolution is parallelized on \glspl{pe}. In the double-buffer implementation these two operations are executed concurrently on \glspl{pe} and \glspl{te} for two different iterations. The runtime results in \Cref{fig:temporal_diagram} correspond to 3$\times$3 2D-convolutional filters applied to 32$\times$16 input frames with 512-element depth.

The \gls{mha} block is present in multiple \gls{che} models~\cite{liu_cevit_2023,wu_aider_2024,wu_darnet_2026}. We implement the version used in \cite{liu_cevit_2023}, the generation of Q, K, and V, the computation of the attention matrix, and the final output-projection step is assigned to \glspl{te}. In particular, we parallelize the computation over the attention heads (H). For the concurrent implementation, we first compute the K-projection, and then overlap the generation of Q and V with the transposition of K. The runtime in \Cref{fig:temporal_diagram} corresponds to four attention heads and Q, K, V matrices of size 128$\times$512.

The results in \Cref{fig:temporal_diagram} show that when operating \glspl{te}, \glspl{pe}, and \gls{dma} concurrently, contentions from simultaneous access to L1 shared memory banks reduce the utilization of the compute and data movement engines. In particular, the average utilization of the \glspl{te}' \glspl{fma} is respectively 67\%, 37\%, and 64\%, for the \gls{fc} layer, the depthwise-separable convolution, and the \gls{mha}. Compared to a purely sequential execution, the runtime is reduced by 16\%, 25\%, and 1.3\%. Notably, these large-dimensional problems fit in the 4MiB L1 of TensorPool, which allows one to overlap large memory transfers and computation, and to solve the data dependencies of \gls{mha} without resorting to L2 transfers.

\begin{figure}[t!]
  \centering
  \includegraphics[width=\columnwidth]{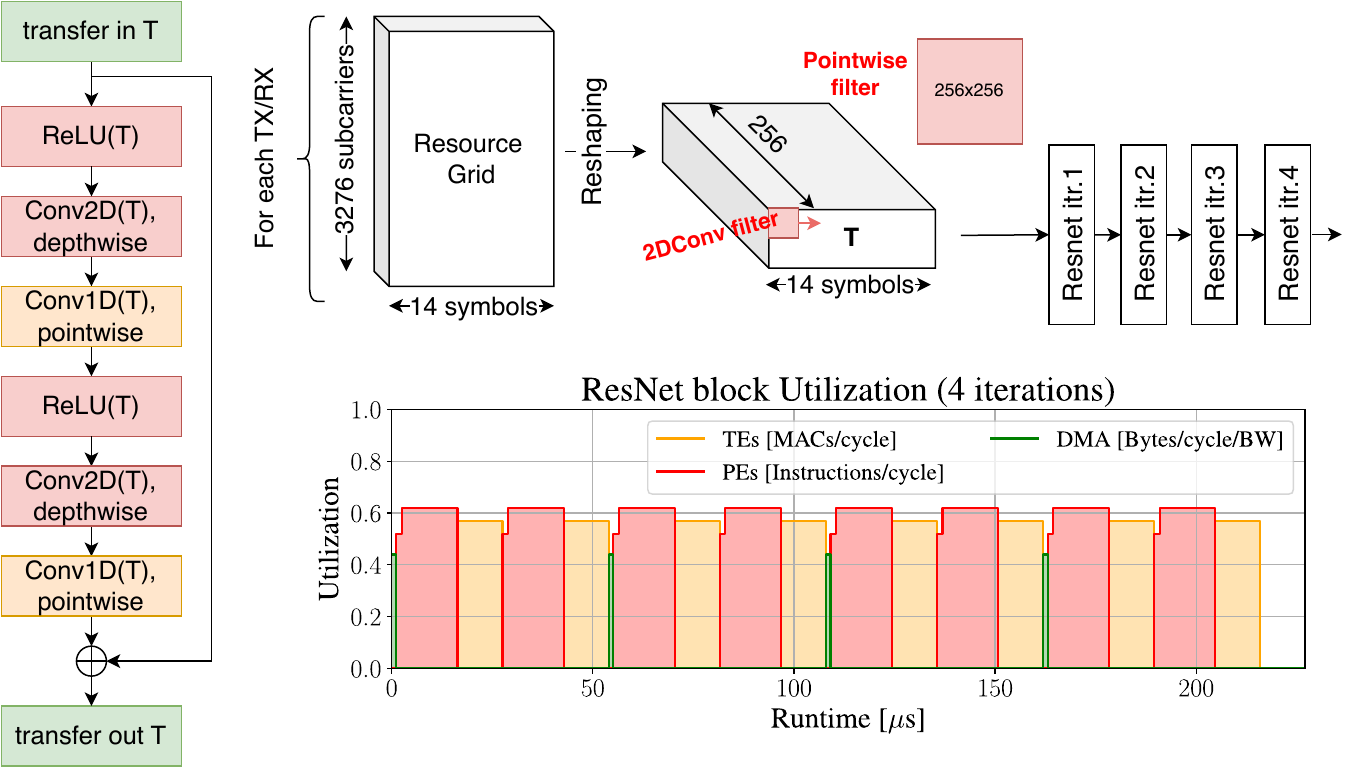}
  \caption{Runtime and utilization of \glspl{te}, \glspl{pe}, \gls{dma}, while executing a ResNet block.}
  \label{fig:resnet_perf}
\end{figure}

\Cref{fig:resnet_perf} schematizes the ResNet blocks used in \cite{wiesmayr_rtnrx_2025,abdollahpour_mdx_2025}. The bulk of \cite{abdollahpour_mdx_2025} consists of four iterations on this block, contributing to 96.7\% of the total runtime.
We consider an input resource-grid with 273~\glspl{prb} (3276~\glspl{re}) and 14 symbols for each transceiver. The resource-grid fits entirely in TensorPool's L1. 
The depthwise convolution and the \gls{relu} execute on \glspl{pe}, the pointwise convolution executes on the \glspl{te}. Assuming 0.9~GHz operation, the time-based utilization trace of \Cref{fig:resnet_perf} shows 0.23~ms runtime, and 57\% \gls{te} utilization on the target problem size. A single TensorPool can address a real-time \gls{mimo} system with up to 4 $N_{TX}\times N_{RX}$.
 
\section{2D Physical Design \& Routing Bottlenecks}
\label{sec:2D_design}

\begin{figure}[t!]
  \centering
  \includegraphics[width=0.85\columnwidth]{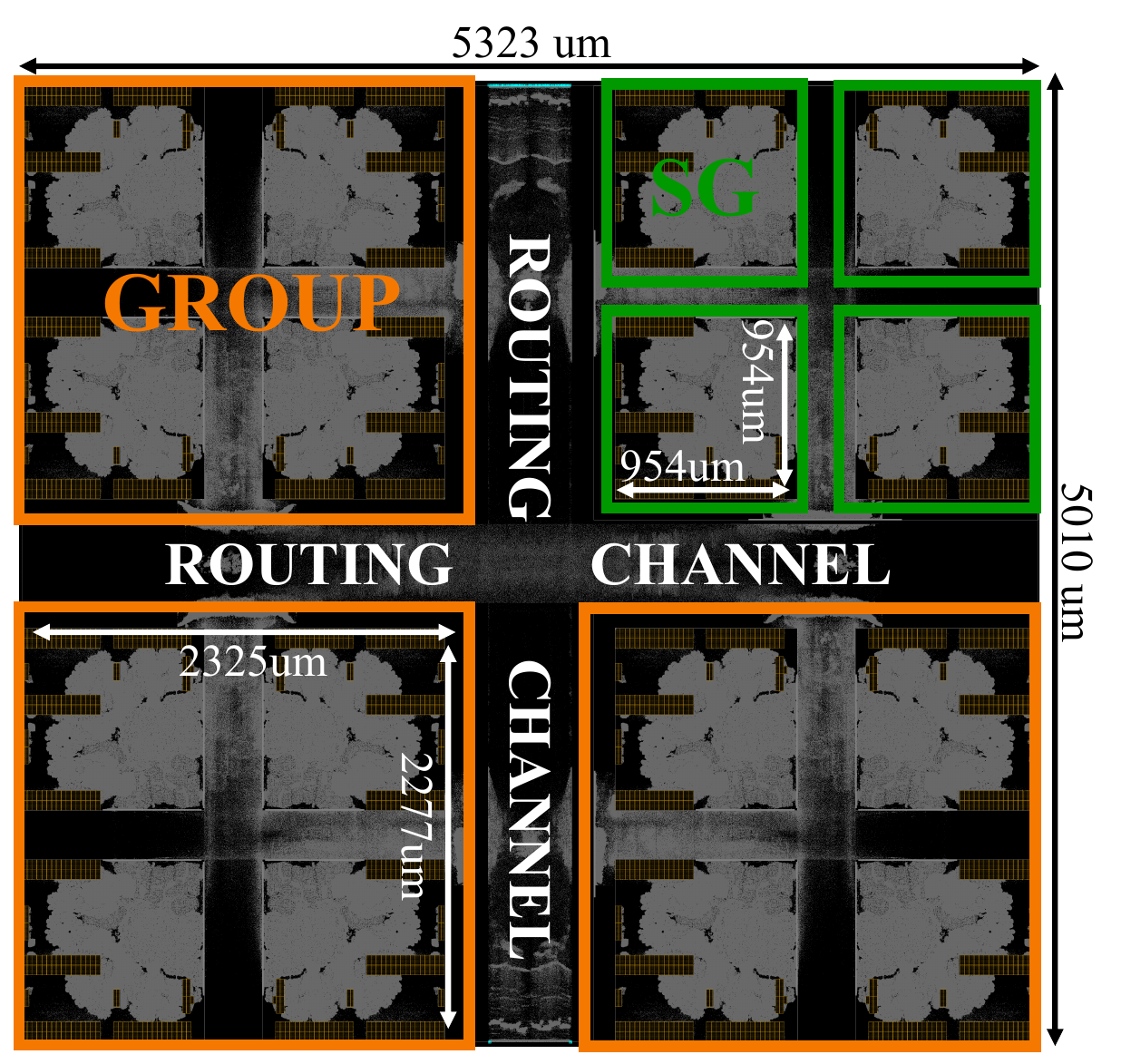}
  \caption{Die snapshot of the TensorPool 2D implementation.}
  \label{fig:die}
\end{figure}

We placed and routed a 2D instance of TensorPool in TSMC-N7 using Synopsys Fusion Compiler 2025.06. We report a screenshot of the placed and routed die in \Cref{fig:die}. The design was assembled hierarchically, bottom-up, starting from the SubGroup, and setting timing constraints according to the position of spill-registers at the hierarchy boundaries.


\begin{figure}[t!]
  \centering
  \includegraphics[width=\columnwidth]{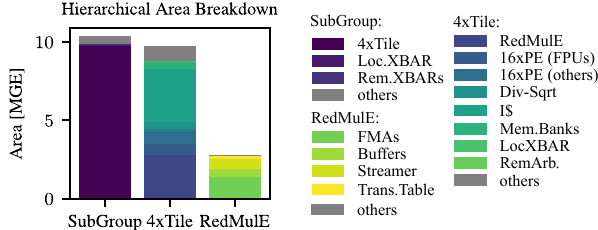}
  \caption{Area Breakdown for the TensorPool SubGroup.}
  \label{fig:area_breakdown}
\end{figure}

\Cref{fig:area_breakdown} reports the area breakdown for a SubGroup instance. The buffers and streamer area is the price to pay for latency tolerance. The X, W, Z buffers, which together occupy 17.6\% of the total \gls{te}'s area, represent a latency-tolerance built-in feature of the \gls{te}: they are also required to feed the engine during the baseline operation in a system with single-cycle access to a 32$\times$ smaller L1-memory than TensorPool's, as presented in \cite{tortorella_redmule_2023}. The streamer's X and W \glspl{rob}, the transactions table, and the Z FIFO, which specifically implement the outstanding burst transaction support, correspond to 31.6\% of the \gls{te}'s area and only 8.5\% of the total SubGroup area. Despite almost 50\% buffering area overhead, each \gls{te} secures peak 1682~MACs/cycle/mm\textsuperscript{2} (\gls{FP16}), while the \glspl{pe}' \glspl{fpu} only provide 752~MACs/cycle/mm\textsuperscript{2} (\gls{FP16}): a 2.23$\times$ improvement in peak compute density over a \gls{pe}-only design.

\begin{table}[b!]
\caption{TensorPool improvement over TeraPool}
\label{tab:comparison}
\resizebox{\columnwidth}{!}{
\begin{tabular}{l r r r}
\toprule
 & TeraPool & \multicolumn{2}{c}{TensorPool} \\
\midrule
Node                                          & 12nm & \multicolumn{2}{c}{7nm} \\
Area (SubGroup) [$\text{mm}^2$]               & 3.0  & \multicolumn{2}{c}{0.9} \\
Area (Group) [$\text{mm}^2$]                  & 17.5 & \multicolumn{2}{c}{5.3} \\
Area (Pool) [$\text{mm}^2$]                   & 81.7 & \multicolumn{2}{c}{26.6} \\
\midrule
\textsuperscript{*}Frequency (TT-25°C) [GHz]  & 0.9  & \multicolumn{2}{c}{0.9} \\
Peak-Performance (TEs) [TFLOPS@\gls{FP16}]        & -    & \multicolumn{2}{c}{7.4} \\
Peak-Performance (TEs + PEs) [TFLOPS@\gls{FP16}]  & 3.7  & \multicolumn{2}{c}{8.4} \\
\midrule
GEMM - Throughput [FP16-MACs/cycle]
& 609
& 3643
& \textcolor{ForestGreen}{6$\times$} \\
GEMM - Performance [TFLOPS@\gls{FP16}]
& 1.10
& 6.62
& \textcolor{ForestGreen}{6$\times$} \\
GEMM - \textsuperscript{\textdagger}Power (TT-25°C,0.75V) [W]
& 6.33  
& 4.32
& \textcolor{ForestGreen}{1.5$\times$} \\
GEMM - Energy Efficiency [TFLOPS@\gls{FP16}/W]
& 0.17
& 1.53
& \textcolor{ForestGreen}{8.8$\times$} \\
GEMM - \textsuperscript{\textdagger}Area Efficiency [TFLOPS@\gls{FP16}/mm\textsuperscript{2}]
& 0.07
& 0.25
& \textcolor{ForestGreen}{6.2$\times$} \\
GEMM - Energy\&Area Efficiency [GFLOPS@\gls{FP16}/W/mm\textsuperscript{2}]
& 6.24
& 57.53
& \textcolor{ForestGreen}{9.1$\times$} \\
\bottomrule
\end{tabular}
}

\begin{tabular}{p{\columnwidth}}
\textsuperscript{*} The critical path of TensorPool is within a \gls{pe}; we compare the two designs at the same target frequency. \\
\textsuperscript{\textdagger} Power \& area technology normalized, multiplying by $(\frac{0.75V}{0.8V})^2$ and $(\frac{7}{12})^2$.
\end{tabular}

\end{table}

\begin{figure}[t!]
  \centering
  \includegraphics[width=\columnwidth]{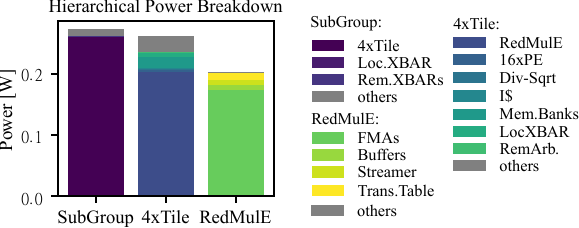}
  \caption{Power Breakdown for the TensorPool SubGroup in the inner loop of a 512$\times$1024$\times$512 \gls{gemm}.}
  \label{fig:power_breakdown}
\end{figure}

We run power simulations of the entire cluster with post placement and routing SubGroup instances in Synopsys PrimeTime 2022.03, and the breakdown for a \gls{gemm} workload is reported in \Cref{fig:power_breakdown}. The SubGroup consumes 0.27~W, 63.7\% of which is in RedMulE's \glspl{fma}, the rest is split among RedMulE's streamer and buffers (11\%), SRAM memory macros (7\%), and interconnects (3.3\%). In \textit{others}, we report the contributions of backend-optimization cells, which were ungrouped in the different hierarchy levels.

\Cref{tab:comparison} is a comparison of TeraPool and TensorPool. TensorPool has 2$\times$ more \glspl{fma} than TeraPool, a homogeneous cluster with 1024 RISC-V \glspl{pe} (2.25$\times$ including \glspl{pe}), and 3$\times$ better utilization during the execution of \gls{gemm}. Therefore, it achieves 3643~MACs/cycle (\gls{FP16}): 6$\times$ more than TeraPool. The higher \gls{fma} utilization, combined with the improvement in peak compute density, concretizes in 6.2$\times$ better area efficiency than TeraPool's (0.25~TFLOPS@\gls{FP16}/mm\textsuperscript{2}). Based on the results of the SubGroup power simulations, TensorPool scores 1.53~TFLOPS@\gls{FP16}/W, 57.53~GFLOPS@\gls{FP16}/W/mm\textsuperscript{2}: respectively 8.8$\times$ and 9.1$\times$ (almost one order of magnitude) higher than TeraPool, thanks to domain specialization.

When assembling the entire cluster bottom-up, the central routing channels must accommodate numerous diagonal connections between Groups. Routing channels occupy 31\% of the Group's area and 21\% of the Pool's area. In 2D-implementation, the Pool is 1.83$\times$ less area-efficient than a SubGroup. The areas' progression in \Cref{tab:comparison} and \Cref{fig:die} show the drop in area efficiency. In the next section, we devise 3D connections between Groups to achieve a better footprint and area efficiency, ensuring continuous performance scalability of the Pool.

\begin{figure}[t!]
  \centering
  \includegraphics[width=\columnwidth]{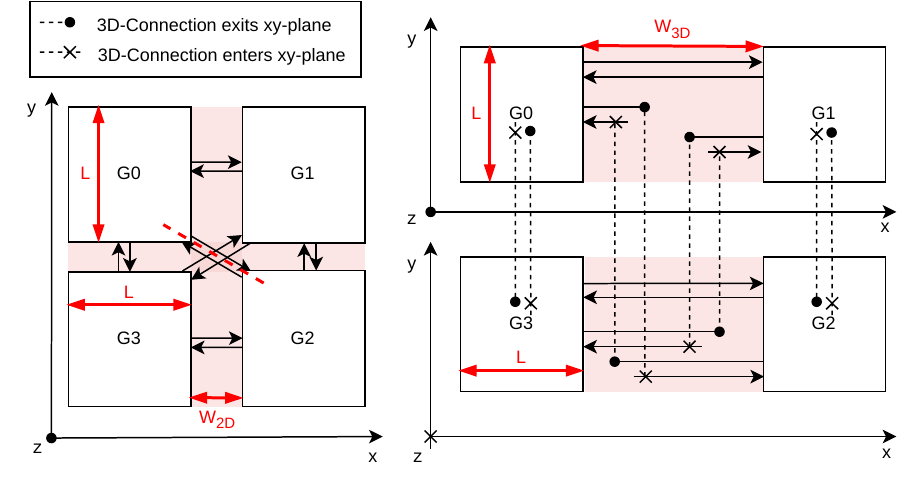}
  \caption{Scheme of the 2D and 3D floorplans. The scratchpad L1-memory connections between Groups are highlighted.}
  \label{fig:floorplans}
\end{figure}

\begin{figure}[t!]
  \centering
  \includegraphics[width=\columnwidth]{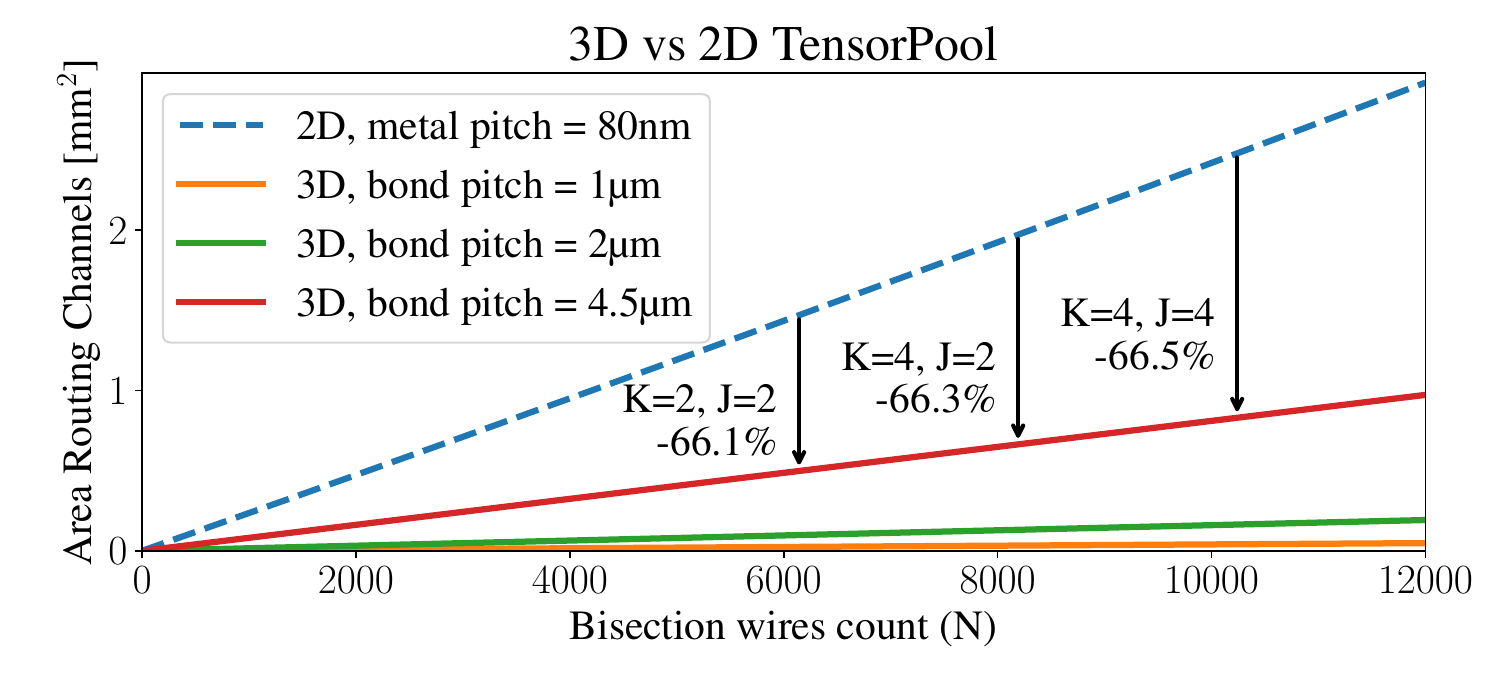}
  \caption{Comparison of the areas of routing channels in a 2D versus a 3D implementation.}
  \label{fig:3d_area}
\end{figure}

\begin{figure}[h!]
  \centering
  \includegraphics[width=0.8\columnwidth]{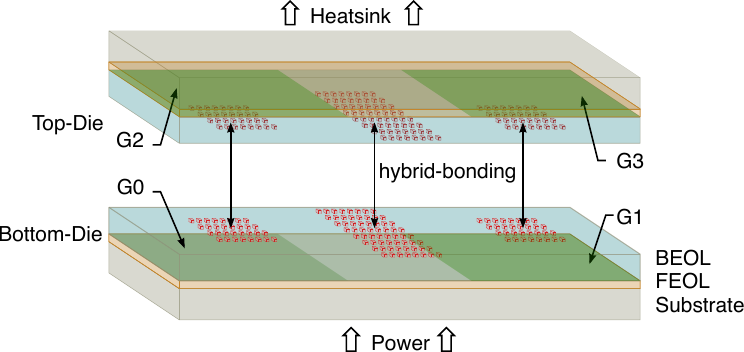}
  \caption{Visualization of 3D-planning for the wafer-to-wafer hybrid-bonded chip. The vertical connections between Groups are either connected directly for Groups sitting on top of each other or routed in the middle channel between Group macros.}
  \label{fig:3d_stack}
\end{figure}

\section{3D Stacked TensorPool}
\label{sec:3D_design}

Previous work on the 3D integration of shared-L1 memory Pools focused on demonstrating the benefits of \gls{mol} partitioning to enlarge the on-chip tightly-coupled shared scratchpad memory size and increase performance and energy efficiency on an L2-memory bound workload~\cite{cavalcante_3dpool_2022}. Further explorations~\cite{das_3dpool_2024, das_3dpool_2025} discussed how to reduce the maximum cores to L1-memory access latency and include an on-chip L2-memory hierarchy, by exploiting up to three tiers of \gls{mol} 3D stacking. In this section, we expand on the discussion carried out by \cite{cavalcante_3dpool_2022}, demonstrating that $\leq5 \mu m$ pitch wafer-to-wafer hybrid-bonds, and abundant \gls{beol} routing resources available in scaled technologies, can be used to reduce the routing congestion at the boundary of the Pool's hierarchies, and the area footprint of routing channels.

\subsection{3D Partitioning}

The four Groups of TensorPool can be split across two dies: the top die contains two of the Groups and routing, the bottom die contains the other two Groups, system registers, the register-based frontend to program \gls{dma} transfers and the top-level AXI-XBAR. In \Cref{fig:floorplans} we report a schematic floorplan with explicit connections between Groups for a single 2D chip and for the desired stacked 3D wafer-to-wafer hybrid-bonded TensorPool. The 3D implementation achieves two main improvements. First, it eliminates one of the routing channels marked with red shading on the 2D floorplan, reducing the routing area. Second, it also eliminates diagonal connections between groups in the limited area at the center of the die. Connections to diagonally opposed Groups can now be spread in the large channel between Groups and routed horizontally in one die, then vertically, crossing the dies' boundary, and finally horizontally again in the other die, reducing crossings and routing congestion. 

The expected reduction in routing channels area can be computed analytically based on metal stack, bond pitch, and geometrical floorplanning information. We compute the 2D-channels area as a function of the $N$ bisection wires crossing the red-dotted line in the figure. Assuming a $p_{2D}$ metal pitch, to accomodate these wires over $N_{metal}$ routing layers, the 2D channel width $W_{2D}$ must be larger than $Np_{2D}/N_{metal}$. For a fixed Group side length $L$:

\begin{equation}
\begin{aligned}
A_{2D} = 4 \times LW_{2D} + x^2 = 4 \times LNp_{2D} + (Np_{2D})^2
\end{aligned}
\end{equation}

In 3D, the central channel must accommodate $2N$ vertical wires, crossing the chip-to-chip boundary. Given the hybrid-bonds pitch $p_{2D}$, and the channel width $W_{3D}$, we compute the area of the central channel:

\begin{equation}
\begin{aligned}
A_{3D} = W_{3D}L = 2N(p_{3D})^2
\end{aligned}
\end{equation}

\Cref{fig:3d_area} represents a comparison between the area of the two channels for $p_{2D}=80nm$, $N_{metal} = 3$ in both the horizontal and vertical routing direction, and for realistic values of the bond pitch. In particular, we highlight the values of $N$ corresponding to the bisection bandwidth for different configurations of K and J, which parametrize the response and request bandwidth of TensorPool's shared L1-memory interconnects. From this high-level analysis of the routing resources available in the vertical direction, thanks to low-pitch hybrid-bonds, a 3D implementation of the TensorPool design with the vertical stacking of Group macros has the potential for up to 66.3\% reduction of the channel area in the chosen K = 4 and J = 2 configuration.

\subsection{3D Performance and Area}
The 3D floorplan for the Group instances is in \Cref{fig:3d_stack}. We envision a stack of two silicon dies with thermal dissipation over the top-die, power delivery from the back side of the bottom-die, and power grid on $M_z$ metal in the bottom-die \gls{beol}~\cite{das_3dpool_2024, wuu_amd_2022}. We demonstrate that the connections between Groups can be routed in the 3D stacked chip without a significant impact on timing caused by hybrid-bonding. To do so, we implement the Groups as black-box instances, applying timing constraints compatible with the post placement and routing results obtained in the 2D design. We assume a fine face-to-face bond pitch of 4.5$\mu$m, with 1$\mu$m $\times$ 1$\mu$m size, 0.5~$\Omega$ resistance, and 1~fF capactance~\cite{lau_hybrid_bonding}. We define the 3D placement of bottom-die, top-die, and bonds as in \Cref{fig:3d_stack} using Synopsys 3DIC Compiler 2025.06. We then complete the placement and routing of the separate die instances in Synopsys Fusion Compiler 2025.06. Finally, we verify the full-stack timing closure in Synopsys 3DIC Compiler.

Each 3D stacked die occupies an 11.47~mm$^2$ area. Compared to the 2D floorplan, the area of the routing channels is reduced by 67\%, from 5.59~mm$^2$ in 2D to 0.91~mm$^2$ on each 3D stacked die. The total wire length of the 3D global routes between the Pool Groups is \SI{20}{\percent} shorter than in 2D, while the two designs exhibit a similar level of routing congestion. This result indicates that exploiting the \textit{Z} direction brings Group macros closer and reduces the wiring distance. For this reason, the 3D design achieves a 2.32$\times$ smaller footprint compared to its 2D counterpart, which is higher than the expected linear 2$\times$ improvement for a 3D design stacked on 2-tiers.

Concerning timing, the longest cross-tier path corresponds to a remote L1-memory access between Groups located at opposite corners of the two-tier stack.
Including the delay of driving buffers and bonding terminals, this route consists of around \SI{120}{\pico\second} at the typical corner (TT, \SI{0.75}{\volt}, \SI{25}{\celsius}), which corresponds to only \SI{10}{\percent} of the clock period. The SubGroup limits the operating frequency of the design.

Although detailed thermal analysis is outside the scope of this work, prior studies suggest that the proposed 3D TensorPool (power density \(30~\mathrm{W/cm^2}\)) can operate within a manageable thermal regime with no frequency scaling. Thanks to double-sided cooling, the peak temperature of a 3D \gls{gpu} and \gls{hbm} stack (respectively, \(64\mathrm{W/cm^2}\) and \(33\mathrm{W/cm^2}\)) can be reduced to the 2.5D baseline~\cite{chen_3dhbm_2025}. Additionally, thermal-scaffolding with polycrystalline diamond heat-conduction paths has enabled reliable operation (\(T_{\max} < 125^\circ\mathrm{C}\)) of a 15-tier heterogeneous stack at up to \(89~\mathrm{W/cm^2/tier}\), suggesting that advanced thermal-management techniques can support the scaling of TensorPool beyond two 3D-stacked tiers~\cite{lyu_diamond_2024}.

In summary, our analysis shows that 3D-integration offers significant benefits and opens a clear opportunity for scaling the domain-specific TensorPool cluster to match the rapidly evolving memory and compute requirements of future AI-Native RANs.

\section{SoA Comparison}
\label{sec:soa}

The academic literature provides several examples of high throughput and high energy efficiency \gls{asic} implementations of classical wireless signal processing algorithms~\cite{Peng_MMSE_ASIC_TCAS_2018,Tang_MMSE_ASIC_JSSCC_2021,Shahabuddin_MIMO_TVLSI_2021,Guo_FFT_ASIC_TVLSI_2023,Yang_FFT_ASIC_TCAS_2023,Zhang_MMSE_ASIC_ISSCC_2024}. Implementing a fixed function, these circuits cannot be repurposed to the emerging AI-Native \gls{ran} workloads. State-of-the-art \gls{asip} solutions for baseband~\cite{Chen_MIMO_TVLSI_2015,Kultala_Lord_TCAS_2019,Fu_DREAM_TCAS_2022,Attari_VectorASIP_TCAS_2022,Castaneda_MMSE_ASIC_ESSCIRC_2022,Chen_DXT501_CoolChips_2022} also have limited reconfigurability, because they implement a small instruction set overspecialized for traditional 5G \gls{phy} processing.  

While academia lacks concrete examples of programmable baseband processors with the flexibility and performance required to migrate to AI-Native \gls{ran}, in \Cref{tab:soa_table}, we compare TensorPool and TensorPool-3D with commercial \gls{gpu} based platforms and tensor accelerators for \gls{ai}-native \gls{ran}. 

The Qualcomm HTA platform~\cite{qualcommHtaRb5} provides attractive performance per MHz and number of \glspl{te}, but offers limited compute capability (2~TOPS), positioning it for lightweight inference tasks rather than full-scale AI-\gls{ran} processing.

Based on their large availability, high performance and programmability, off-the-shelf \gls{gpu} platforms have been proposed for the acceleration of equalization and decoding functions in 5G base-stations~\cite{tarver_gpu_2021,wu_gpu_2014,li_gpu_2015}. More recently, \cite{lin_airanempowering_2025, kelkar_aerial_2021} identify data-center \glspl{gpu} as best candidates to address the emerging AI-Native \gls{ran} workloads. The Aerial \gls{ran} systems~\cite{nvidiaAerialRanComputerProDatasheet,nvidiaRtxProBlackwellArchV10,velayutham2024aerialran1,nvidiaRtxBlackwellArchV11,atri2025arccompact,nvidiaAdaLovelaceArchV11} are the commercial off-the-shelf products for AI-Native \gls{ran} with the highest tensor throughput in \Cref{tab:soa_table}. They reach up to 503.8~TOPS, by scaling-out the \gls{sm}, a modular shared-L1 compute cluster with 128~KiB of L1-memory, 128 \glspl{pe} and 4 \glspl{te}. 

Base-station hardware architectures should be co-designed and tailored to the specific compute and memory demands of deeply optimized models for the edge, maximizing energy efficiency while preserving flexibility. In \glspl{gpu}, high performance comes at the cost of substantial power consumption. Although power measurements on \glspl{gpu} running 6G workloads have not yet been disclosed by open literature, the typical power reported for \cite{nvidiaAerialRanComputerProDatasheet,nvidiaRtxProBlackwellArchV10,velayutham2024aerialran1,nvidiaRtxBlackwellArchV11} is up to 600~W, significantly higher than TensorPool's. The power densities at peak utilization are within the same order of magnitude: 0.8~\si{W/mm\textsuperscript{2}} on the \gls{gpu} die (based on available measurements~\cite{nvidiaRtxProBlackwellArchV10,nvidiaRtxBlackwellArchV11}, including on-chip L2 cache and memory controllers), compared with 0.18~\si{W/mm\textsuperscript{2}} for TensorPool. Therefore, \cite{nvidiaAerialRanComputerProDatasheet,nvidiaRtxProBlackwellArchV10,velayutham2024aerialran1,nvidiaRtxBlackwellArchV11} \glspl{gpu} are well-suited for centralized deployments~\cite{kelkar_aerial_2021}, but not for the energy-constrained base-station remote units.

The Aerial RAN Compact design \cite{atri2025arccompact,nvidiaAdaLovelaceArchV11} shows lower peak power consumption, compatible with base-station deployment. However, the L1 shared-memory of its \glspl{sm} has limited size: it does not fit the real-imaginary complex \gls{FP16} data of a full resource-grid with 273~\glspl{prb}, 14 symbols, and a single transceiver (180~KiB). TensorPool has 32$\times$ larger L1 memory than a \gls{sm}, it fits 273~\glspl{prb} and 14 symbols for up to 22 transceivers, promoting data-reuse and reducing data transfers to higher memory hierarchies. Compared to a \gls{sm}, where the L1 memory is shared by 4 \glspl{te} only, the 16 \glspl{te} of TensorPool deliver 4.76$\times$ higher performance, guaranteeing real-time end-to-end processing of an \gls{ofdma} \gls{nn}-model for up to 4 transceivers.

Additionally, TensorPool-3D achieves a comparable area efficiency to all the Aerial designs (288~GOPS/mm$^2$), once taking into account the improvement offered by technology scaling. Moreover, it demonstrates 1.16$\times$ higher area-efficiency than its 2D counterpart and a superlinear 2.32$\times$ footprint reduction, highlighting the benefits of a compact 3D-design and opening further scale-up opportunities.


\begin{table*}
\centering
\caption{Comparison of tensor-accelerated platforms for AI-Native RAN.}
\label{tab:soa_table}
\footnotesize
\setlength{\tabcolsep}{4pt}
\renewcommand{\arraystretch}{1.1}
\begin{adjustbox}{width=\textwidth}
\begin{tabular}{lccccccc}
\toprule
& \shortstack[c]{Aerial RAN\\Computer Pro\cite{nvidiaAerialRanComputerProDatasheet}}
& \shortstack[c]{Aerial RAN\\Computer-1\cite{velayutham2024aerialran1}}
& \shortstack[c]{Aerial RAN\\Compact\cite{atri2025arccompact}}
& \cite{qualcommHtaRb5}
& TensorPool
& TensorPool-3D \\
\midrule

Architecture
& \shortstack[c]{NVIDIA Blackwell\\ RTX PRO 6000 GPU~\cite{nvidiaRtxProBlackwellArchV10}}
& \shortstack[c]{NVIDIA Blackwell\\ RTX 5090 GPU~\cite{nvidiaRtxBlackwellArchV11}}
& \shortstack[c]{NVIDIA L4\\ Tensor-Core GPU~\cite{nvidiaAdaLovelaceArchV11}}
& \shortstack[c]{Qualcomm Hexagon\\ HTA230}
& -
& - \\

Number L1-Clusters
& 188 (SM)
& 170 (SM)
& 60 (SM)
& 1
& 1
& 1\\

L1-size
& $128\,\mathrm{KiB}$
& $128\,\mathrm{KiB}$
& $128\,\mathrm{KiB}$
& $128\,\mathrm{KiB}$
& 4\,MiB
& 4\,MiB \\

Number TEs
& 752 $(\mathrm{4}/ \mathrm{SM})$
& 680 $(\mathrm{4}/ \mathrm{SM})$
& 240 $(\mathrm{4}/ \mathrm{SM})$
& 2
& 16
& 16 \\

Number PEs
& 24064 $(\mathrm{128}/ \mathrm{SM})$
& 6144 $(\mathrm{128}/ \mathrm{SM})$
& 7424 $(\mathrm{128}/ \mathrm{SM})$
& -
& 256
& 256 \\

Tech.
& 4\,nm
& 4\,nm
& 4\,nm
& -
& 7\,nm
& 7\,nm \\

$f$ [MHz]
& \num{2617}
& \num{2407}
& \num{2040}
& \num{1000}
& \num{900}
& \num{900} \\

Precision
& FP16
& FP16
& FP16
& Fixed-point 16
& FP16
& FP16 \\

\shortstack[c]{Area [mm$^2$]\\ \quad}
& \shortstack[c]{\num{750}\\ \quad}
& \shortstack[c]{\num{750}\\ \quad}
& \shortstack[c]{\num{294}\\ \quad}
& \shortstack[c]{-\\ \quad}
& \shortstack[c]{\num{26.65}\\ \quad}
& \shortstack[c]{\num{22.94}\\ (\num{11.47} footprint)} \\

Area L1-Cluster [mm$^2$]
& \num{1.7}\textsuperscript{*}
& \num{1.7}\textsuperscript{*}
& \num{1.7}\textsuperscript{*}
& -
& \num{26.65}
& \num{22.94}\\

Power [W]
& \num{600}
& \num{575}
& \num{72}
& -
& \num{4.32}
& \num{4.32} \\

GOPS (TEs)
& \num{503800}
& \num{419000}
& \num{121000}
& \num{2000}
& \num{6623}
& \num{6623} \\

GOPS (TEs) / Number L1-Clusters
& \num{2680} - 1440\textsuperscript{**}
& \num{2465} - 1440\textsuperscript{**}
& \num{2017} - 1390\textsuperscript{**}
& \num{2000}
& \num{6623}
& \num{6623} \\

GOPS (TEs) / Area L1-Cluster [mm$^2$]
& \num{1576} - 277\textsuperscript{\textdagger} 
& \num{1450} - 277\textsuperscript{\textdagger} 
& \num{1190} - 267\textsuperscript{\textdagger} 
& -
& \num{249}
& \num{288} \\

\bottomrule
\end{tabular}
\end{adjustbox}

\vspace{5pt}
\begin{tabular}{p{\textwidth}}
\textsuperscript{*} Area of the \gls{sm}, based on~\cite{ad102_die, b202_die}.\\
\textsuperscript{**} Frequency normalized, based on NVIDIA's A100 (1410~MHz), fabricated in the same technology node of TensorPool (N7)~\cite{nvidia_a100} .\\
\textsuperscript{\textdagger} Area normalized by $(7/4)^2$ to account for technology scaling.\\
\end{tabular}

\end{table*}

\section{Conclusion}
\label{sec:conclusion}

In this work, we presented TensorPool, a many-core cluster with 256 lightweight \glspl{pe} accelerated by 16 256-\glspl{fma} \glspl{te}. TensorPool's heterogeneous design combines \glspl{te} and \glspl{pe} to target the intensive tensor computations in AI-\gls{phy}, at a power consumption compatible with edge basestation deployments. The low-latency interconnect, the novel \glspl{te} memory interface and the streamlined parallelization scheme ensure peak 89\% parallel \gls{fma}-utilization on \gls{gemm}: 3$\times$ the \gls{fpu}-utilization of TeraPool, a shared-memory Pool with \glspl{pe} only. 

Delivering 6.62~TOPS performance on AI computations, TensorPool meets the requirements for real-time processing of AI-Native \gls{ran} baseband models. A placed and routed instance of TensorPool in TSMC's N7 achieves 57.53~GFLOPS@FP16/W/mm\textsuperscript{2}, 9.1$\times$ higher compared to the non-accelerated TeraPool, thanks to better utilization of the compute units. 

The 3D-stacking of TensorPool Groups allows to unfold the large number of connections required to feed the bandwidth of \glspl{te}, and provides superlinear 2.32$\times$ footprint reduction without frequency degradation compared to the 2D implementation. Combined with power consumption well within the specifications (4.3~W), this result opens opportunities for further scaling up the performance of 3D shared L1-memory Pools for AI-Native baseband processing.

\section*{Acknowledgments}
This work was supported by Huawei Sweden AB.

\bibliographystyle{IEEEtran}
\bibliography{REFERENCES.bib}

\vfill

\end{document}